\documentclass[a4paper]{report}
\pdfoutput=1

\usepackage[margin=2.5cm]{geometry}
\usepackage[utf8]{inputenc}
\usepackage[T1]{fontenc}
\usepackage[english]{babel}

\usepackage[pdftex,breaklinks=true,bookmarks=true,pdfborder={0 0 0},colorlinks=false]{hyperref}

\usepackage{array}
\usepackage{amssymb}
\usepackage{MnSymbol}

\usepackage{listings}
\lstdefinelanguage{sl}{alsolanguage=C,morekeywords={sl_def,sl_enddef,sl_index,sl_parm,sl_getp,sl_setp,__asm__,sl_glparm,sl__static,sl_shparm,__volatile__,__typeof__,sl_create,sl_createsync,sl_glarg,sl_sharg,sl_geta,sl_sync,sl__exclusive,sl__static,sl_detach,sl_decl,sl_typedef_fptr,sl_glfparm,sl_glfarg,sl_shfparm,sl_shfarg,inline,sl_seta}}

\usepackage{cleveref}
\usepackage{graphicx}

\newcommand{\ie}{i.e.~}
\newcommand{\eg}{e.g.~}
\newcommand{\cf}{cf.~}
\newcommand{\vs}{vs.~}

\usepackage{enumitem}
\setlist{nolistsep}

\usepackage{float}
\floatstyle{ruled}
\newfloat{sidenote}{b}{sn}[chapter]
\floatname{sidenote}{Side note}
\crefname{sidenote}{side note}{side notes}

\usepackage{framed}
\usepackage[usenames]{color}

\definecolor{shadecolor}{gray}{0.9}
\newenvironment{summary}{%
\-

\phantomsection
\noindent
\begin{minipage}{\textwidth}
\begin{snugshade}
\section*{Chapter summary}%
}{\end{snugshade}
\end{minipage}
}

\begin{document}

\author{Raphael ‘kena’ Poss\\University of Amsterdam, The Netherlands\\\url{r.poss@uva.nl}}
\title{On whether and how \\ D-RISC and Microgrids can be kept relevant \\
(self-assessment report)}

\maketitle

\begin{abstract}
This report lays flat my personal views on D-RISC and Microgrids as of
March 2013. It reflects the opinions and insights that I have gained
from working on this project during the period 2008-2013.  This report
is structed in two parts: deconstruction and reconstruction. In the
deconstruction phase, I review what I believe are the fundamental
motivation and goals of the D-RISC/Microgrids enterprise, and identify
what I judge are shortcomings: that the project did not
deliver on its expectations, that fundamental questions are left
unanswered, and that its original motivation may not even be relevant
in scientific research any more in this day and age. In the
reconstruction phase, I start by identifying the merits of the current
D-RISC/Microgrids technology and know-how taken at face value,
re-motivate its existence from a different angle, and suggest new,
relevant research questions that could justify continued scientific
investment.
\end{abstract}

\setcounter{tocdepth}{1}
\tableofcontents

\clearpage

\section*{Disclaimer}
{\itshape

The arguments presented hereafter are my own, and thus may not be shared
by my colleagues or work partners. To my knowledge, at the time of
this writing there is no acknowledgement or consensus around the
D-RISC/Microgrids enterprise, other than my own experience and
impressions, that give credit to the perspective presented here.
}

\section*{Purpose and rationale}

This report lays flat my personal views on D-RISC and Microgrids as of
March 2013. It reflects the opinions and insights that I have gained
from working on this project during the period 2008-2013. 

The origin of this report is a case of cognitive dissonance. On the
one hand, using critical thought against my “achievements” of the past
few years is causing a growing discomfort, discontent and
disappointment at the way the design and implementation of
D-RISC/Microgrids have been carried out so far, both by myself and my
colleagues. On the other hand, my optimism combined with an unusual
combination of curiosity and fascination for theoretical computer
science is sustaining a belief that despite its flaws, the project has
produced an interesting conceptual framework which deserves further
investigation at least by academics and teachers. Only by resolving
this cognitive dissonance can I satisfy myself that my continued work
in this area is compatible with my aspirations as a scientific
researcher.  By writing this report, I hope I can resolve this
dissonance by externalizing both sides and construct rationally their
resolution.

\clearpage
\section*{Executive summary}

\paragraph{Shortcomings in the research results so far.}
The D-RISC/Microgrids project was purportedly intended to solve major issues in
micro-architecture research, related to scalability in performance and
efficiency in general-purpose microprocessors. 
The strategy to solve these issues was to implement a
combination of dataflow scheduling with hardware support for thread
concurrency management within and across cores on chip. 
Implementation
was carried out, but the results are inconclusive. On the one hand, the proposed
hardware does indeed provide higher performance and efficiency in
regular, data-parallel computation kernels. On the other hand, no evidence has yet
been produced that the proposed hardware benefits larger applications
with more irregular workloads. 
Power efficiency and intelligent resource
management was regularly advertised but not actively researched.
Effort has been invested into widening the scope of the technology
towards applications and industrial relevance, but these applications
have not yet materialized.

\paragraph{Shortcomings in methodology.}
Research on D-RISC/Microgrids is not following the scientific method.
It is instead currently carried out as an engineering enterprise, but without clear
technology ouputs and without identifying
its potential applications. Its relevance in a university research group is thus questionable.

\paragraph{Obstacles to further progress.}
The D-RISC/Microgrids project has the ambitious aim to produce a
general-purpose processor chip able to disrupt the current
state-of-the-art. However, the limited human resources dedicated to
the project are insufficient to reach this aim in isolation.
The expansion of the research group to a community of users and and research
partners is blocked by a fundamental lack of compatibility with existing operating
systems and application software. This lack of
compatibility is not properly justified, neither by practical nor
theoretical reasons.
Meanwhile, the scientific effort to test the hypotheses that underly
the D-RISC/Microgrids project is poorly directed, and not
enough attention has been given to negative results that
invalidate these hypotheses.
Finally, the multi-core research field is nowadays much
more crowded than it was ten years ago, yet the research on D-RISC/Microgrids
does not acknowledge its competition nor attempts to differentiate
its contributions from the state of the art.

\paragraph{Actual contributions.}
 The research has produced interesting discussions
  that challenge some tacit assumptions of the research community, 
  experimental results that can be reused by future work, improvements
  to partner technologies and new simulation techniques.
 Most of the software designed and implemented during the research
  can be reused by third parties, and not only for research
  directly related to D-RISC/Microgrids.
 The intellectual framework educates practitioners to think about
  two general separations of concerns, namely concurrency \vs parallelism and
  using memory for storage \vs synchronization.

\paragraph{Individual architectural features.}
The D-RISC core combines features found in other processors, such as a RISC pipeline
  and hardware multithreading, with custom features (\eg its TMU) and optimizations
  to the conventional features (\eg switch annotations for the HMT scheduler).
Some architectural optimizations found in D-RISC/Microgrids could be reused 
  with other processors, for example switch annotations and bulk coherency in the memory network.
The key feature of D-RISC/Microgrids, namely its TMU and inter-TMU control NoC, 
  does not depend on the other features specific to D-RISC and could be potentially reused with other processors.

\paragraph{Follow-up strategies.}
I can see three follow-up strategies for new investments around
  D-RISC/Microgrids: exploitation, \ie apply the technology produced
  so far to other uses than research; salvaging and opening the
  technology, \ie extracting individual features from the
  D-RISC/Microgrids design and evaluating them as extensions of
  existing processors; and distillation of the main ideas in the realm
  of fundamental computer science.
 Ongoing research towards doctoral theses should be careful
  to rephrase research questions in the light of our recent shared
 understanding of the project's issues.

\chapter{Background}

\section{Scalability in general-purpose processors}

Assuming continued demand for computers where \emph{new functions can
  be defined by the end-user by writing and using their own software}
(cf. \cref{sn:gpdemand}), the question remains of how to make
general-purpose processors that are both fast (operations/second) and
efficient (operations/watt). However, conventional approaches on
silicon seem to have reached ``walls'' on both fronts around year
2000~\cite{ronen.01.ieee}.

Since technology progress still delivers increasingly more transistors
per chip (Moore's law), the trend has become to glue individual
processors together on the same chip, i.e. design ``multi-cores.''
The issue with this is that software is mostly written using
sequential algorithms: introducing hardware parallelism (multiple
processors) immediately raises the question of how to introduce
explicit concurrency in software. Software concurrency is hard and
both hardware architects and programming language designers have been
making only baby steps since 2000.

\begin{sidenote}
\caption{On the continued demand for general-purpose
  computers.}\label{sn:gpdemand} One needs to accept the premise that
general-purpose computers are highly desirable and that the future of
computing hangs on their continued development to fully value the
remainder of this report. I have explained my own reasons to accept
this premise in \cite[sect.~1.2--1.4]{poss.12}; in short, I propose
that general-purpose computers are, like “stem cells,” necessary to
the continuation of computer science. I also propose they are
essential to the democratic freedom of any citizen to create their own
tools (in software) in this numeric age.

Meanwhile, I also acknowledge that it is not the role of mere
academicians to decide “what people really want.” If market, fashion
and politics determine that science should research instead all sorts of
maximally efficient special-purpose computing devices, the deconstruction
phase of this report would even be easier: there would be simply
no place at all for D-RISC and Microgrids.
\end{sidenote}

\section{D-RISC and Microgrids: what has been done}

The \emph{Microgrid} many-core architecture is a research project at
the University of Amsterdam, which investigates whether concurrency
management (thread scheduling, synchronization, and inter-thread
communication) traditionally under control of software operating
systems can be accelerated in hardware to obtain higher efficiency and
performance. Microgrids are clusters of a simple RISC core design
called D-RISC~\cite{bolychevsky.96.ieee}; each D-RISC core supports
hardware multi-threading (HMT) using a dataflow scheduler, and is also equipped with a hardware Thread
Management Unit (TMU) which can coordinate with neighbouring TMUs for
automatic thread and data distribution (cf. \cref{sn:tmu}). In short:

\begin{center}
\bfseries
D-RISC = simple RISC + dataflow HMT scheduler + TMU $-$ interrupt management

Microgrid = n$\times$D-RISC + TMU-to-TMU NoC + custom cache/memory protocol
\end{center}

\begin{sidenote}
\caption{Details about D-RISC/Microgrids architecture}\label{sn:tmu}
The rest of the text assumes passing familiarity with the D-RISC
and Microgrids architecture, as presented in chapters 3 and 4 of~\cite{poss.12}.
\end{sidenote}

Prior to 2007, research on D-RISC and the Microgrid was focused on
programmability issues and carried out with high-level simulators:
both using traditional software multithreading and an API to emulate
the TMU services~\cite{tol.09.jsa}, and using a custom functional ISA
emulator~\cite{bousias.06.cj}. As the initial phases of the D-RISC and
Microgrid design were encouraging~\cite{bell.06.jpp,bousias.09.jsa},
the EU-funded project Apple-CORE (2008-2011) was started to study its
implementability in a system, including a full vertical tooling stack
from an FPGA implementation up to benchmarks in higher-level
programming languages.

The outcome of the Apple-CORE project is summarized
in~\cite{poss.12.dsd,poss.12}: the D-RISC core was implemented on FPGA
as UTLEON3~\cite{danek.12}, a model of Microgrids was implemented in
MGSim, software tooling was delivered to program
Microgrids~\cite{saougkos.11,poss.12.sl,grelck.09.cpc}, and D-RISC and
Microgrids were confirmed using both UTLEON3 and MGSim to deliver
higher performance and efficiency for \emph{some of the selected
  benchmarks}.

\section{D-RISC and Microgrids: what is going on}

At the time of this writing, research in this area continues on two
fronts.  An industry-backed project has funded more effort towards
tailoring D-RISC for real-time embedded systems, by adding
priority-based scheduling and fault tolerance. Next to this, four
doctoral candidates are planning to defend their thesis on extensions
and improvements to D-RISC and Microgrids, and simulations thereof.
The Microgrid-related technology produced so far is also used for
graduate and undergraduate education in computer architecture and
compiler construction.

\part{Deconstruction}\label{part:decons}

\chapter{Outcomes \vs original intents: a retrospective}
\label{chap:outcomes}

The research on these topics was funded based on multiple research
proposals and statements of intent over time. In this section, I
explore \emph{published} motivations, where the scientific community has
agreed via the publication approval process that the motivations and
justifications were worthwhile.
 
Although most scientific projects are expected to diverge from their
original intents, published objectives reveal the general motivation
that drives the effort. Note that I do not focus here on
``result-oriented'' papers that report factually on research outcomes;
only on those publications that made statements of intent and
motivation before the corresponding work was carried out.

\section{I. Bell, N. Hasasneh \& C. Jesshope, JPP 2006}

Here we explore the contributions originally advertised in~\cite{bell.06.jpp}.

\subsection{Summary of the article}

Abstract:
\begin{quote}
\itshape Chip multiprocessors (CMPs) hold great promise for achieving
scalability in future systems. Microthreaded CMPs add a means of
exploiting legacy code in such systems. Using this model, compilers
generate parametric concurrency from sequential source code, which
can be used to optimise a range of operational parameters such as
power and performance over many orders of magnitude, given a scalable
implementation. This paper shows scalability in performance, power
and most importantly, in silicon implementation, the main contribution
of this paper. The microthread model requires dynamic register
allocation and a hardware scheduler, which must support hundreds of
microthreads per processor. The scheduler must support thread
creation, context switching and thread rescheduling on every machine
cycle to fully support this model, which is a significant
challenge. Scalable implementations of such support structures are
given and the feasibility of large-scale CMPs is investigated by
giving detailed area estimate of these structures.
\end{quote}

Problem statement:
\begin{quote}\itshape
In general, there are only a few requirements for the design of
efficient and powerful general-purpose CMPs, these are: scalability of
performance, area and power with issue width, and programmability from
legacy sequential code. Issue width is defined here as the number of
instructions issued on chip simultaneously, whether in a single
processor or in multiple processors and no distinction is made
here. To meet these requirements a number of problems must be solved,
including the extraction of ILP from legacy code, managing locality,
minimising global communication, latency tolerance, power-efficient
instruction execution strategies (i.e. avoiding speculation),
effective power management, workload balancing, and finally, the
decoupling of remote and local activity to allow for an asynchronous
composition of synchronous processors. Most CMPs address only some of
these issues as they attempt to reuse elements of existing processor
designs, ignoring the fact that these are suitable only for chips with
relatively few cores.
\end{quote}

Proposed main contribution:
\begin{quote}
\bfseries\itshape
In this paper a CMP is evaluated, that is based on microthreading, which addresses either directly or indirectly, all of the above issues and, theoretically, provides the ability to scale systems to very large number of processors.
\end{quote}

\subsection{Analysis}

The ``CMP'' in the paper's text refers to the D-RISC/Microgrid technology. Does this paper, and all the research
since then (8 years) support the claim that it solves ``all of the above issues''?
We list here how the technology addresses the issues in decreasing order of success.

\paragraph{Extraction of ILP from legacy code:} this was successful. ILP is extracted implicitly by D-RISC's dataflow scheduler, although
the ILP width is limited by the number of active threads and the number of registers per thread, because the reordering
information is stored in registers.

\paragraph{Decoupling of remote and local activity:} this is mostly successful, insofar the D-RISC's TMU control protocol
has different primitives to spawn work locally and remotely. 

\paragraph{Scalability of performance, area and power with issue width:} each D-RISC core uses a single-issue pipeline,
so this claim states that performance, area and power scales with the number of cores. The research has indeed
shown this to be true for large, regular, data-parallel computation kernels, but the picture is not so clear for
small or more heterogeneous workloads because of the inter-core latencies in the concurrency management protocol itself
and on-chip network contention due to cache coherency protocols.

\paragraph{Power-efficient instruction execution strategies:} this is only partly successful; with one thread
the execution is not too power efficient compared to traditional
single-issue, in-order designs, and efficiency only increases with the number of threads
active. The problem is that operand availability is only tested at the
read stage of the pipeline. When there are multiple threads,
annotations at the fetch stage prevent ``potentially suspending''
instructions from entering the pipeline, but if only one thread is
active, the instructions with missing operands will still enter the
pipeline, create a bubble after the read stage, and subsequently need
to be rescheduled. This is a form of speculation, thus inherently less
power-efficient than the traditional approach to stall the pipeline during
issue.

\paragraph{Latency tolerance:} this is only partly successful. The latency of intra-core operations is tolerated by intra-thread ILP, but then the same is possible with conventional barrel processors or out-of-order execution\footnote{Albeit possibly at a larger area and power budget. However, the actual area and power requirement of D-RISC are yet to be evaluated.}. Longer latencies can be tolerated as long as there are sufficient threads active on the core to interleave with a waiting thread. If most active threads are busy communicating, then latency tolerance is highly dependent on a full end-to-end support for split-phase transactions, \ie the rest of the system must support
a large number of in-flight transactions. In practice, the caches and
external memory interfaces become a bottleneck, and to this day no clear solution has emerged on this front. 

\paragraph{Minimising global communication:} this is only mildly successful. For synchronization and concurrency management, global communication is limited by constraining the TMU's automatic workload distribution to adjacent cores only. The responsibility of choosing an area of the chip where to start the distribution is left to an hypothetical resource manager, not yet researched/implemented. Next to this, one should also consider memory communication. Here all results so far use memory protocols that incur global communication for coherency. No results show yet that global memory communication has been minimized.

\paragraph{Workload balancing:} this is only very mildly successful. D-RISC's TMU can automatically spread a batch of threads to multiple cores using an even N/P distribution, but this is the only form of distribution supported. Due to bulk reuse and synchronization, this simple distribution causes irrecoverable imbalances as soon as the batch is heterogeneous. 

\paragraph{Managing locality:} this is not achieved, insofar that the data used by instructions is invisible to D-RISC's ``intelligence'' (its hardware TMU), and the memory and core networks are not topologically congruent. The software has to negociate locality of code and data explicitly with knowledge of the chip's layout.

\paragraph{Programmability from legacy sequential code:} to this date, most existing sequential code cannot
be reused as-is with D-RISC/Microgrids, because of incomplete support
for operating system services, \cf also \cref{sec:incompat}.

\paragraph{Effective power management:} to this date, power management has not been explored.
 
\section{NWO Microgrids: 2006-2010}

\subsection{Research question}

The project NWO Microgrids was funded by the Dutch government based on
the following research question:

\begin{quote} \itshape
Is it possible, through the introduction of simple and explicit concurrency controls, to develop a systematic approach to:
\begin{enumerate}
\item incrementally designing new processor architectures (i.e. based on an existing ISA and infrastructure);
\item dynamically managing and optimising the available resources for a variety of goals such as performance, power and reliability (i.e. resulting in autonomous and self-adaptive microgrids);
\item formally defining the architectures' execution properties; 
\item incrementally developing the architectures' infrastructure (i.e. simulators, compilers, binary-to-binary translators and even silicon intellectual property); 
\end{enumerate}
all within the context of ten to fifteen years of silicon-technology scaling (i.e. maintaining scalability over a thousand fold increase in chip density)?
\end{quote}

Note that this is a yes/no question. 

To answer ``yes,'' it suffices to propose at least one set of ``simple
and explicit concurrency controls'' and a corresponding ``systematic
approach'' that delivers on the four other points. Alternatively,
``yes'' can also be given, perhaps less satisfactorily, if a
theoretical analysis indirectly merely proves the systematic approach 
exists (``it is possible to develop it'') without actually developing
it.  To answer ``no,'' in contrast, it is necessary to demonstrate
that there cannot exist any set of ``simple and explicit concurrency
controls'' which makes a ``systematic approach'' possible.

Strategically, this question suggests its own answer:
\begin{itemize}
\item a ``no'' answer would be a formidable theoretical endeavour, likely very
difficult to obtain (possibly impossible within the proposed 10-15
years time frame);
\item a ``yes'' answer based on theoretical proof of existence would be equally difficult;
\item therefore, the question suggests a ``yes'' is expected, hinged
  on the ability of the researchers to use the features of an existing
  architecture as their candidate ``simple and explicit concurrency
  controls,'' and consequently show that a corresponding ``systematic
  approach'' delivers on the 4 other points.
\end{itemize}

In short, NWO Microgrids's ``declaration of scientific intent'' can be reformulated as follows:

\begin{quote} \itshape
We will show that D-RISC/Microgrids make it possible to develop a systematic approach to:
\begin{enumerate}
\item incrementally designing new processor architectures (i.e. based on an existing ISA and infrastructure);
\item dynamically managing and optimising the available resources for a variety of goals such as performance, power and reliability (i.e. resulting in autonomous and self-adaptive microgrids);
\item formally defining the architectures' execution properties; 
\item incrementally developing the architectures' infrastructure (i.e. simulators, compilers, binary-to-binary translators and even silicon intellectual property); 
\end{enumerate}
all within the context of ten to fifteen years of silicon-technology scaling.
\end{quote}

\subsection{Outcome \vs expectations}

Did NWO Microgrids deliver on its self-set expectations?

\paragraph{Did NWO Microgrids deliver a ``systematic approach'' with the desired properties?} No, the design of D-RISC/Microgrids
was instead carried out in an ad-hoc fashion, with multiple phases of
trial-and-error and backtracking. I highlight an ethical issue here,
because on the one hand the Dutch NWO funded a project on the
assumption that its outcome would be a \emph{systematic approach
  (method)} that could be reused with different architectures, and on
the other hand the research team knew well in
advance that systematization would not be studied.

For the sake of deconstruction, let us however stretch the word
``systematic approach'' to encompass ``the process of designing and
building D-RISC/Microgrids.'' Does this extended definition match the
other required properties?

\paragraph{Did the process of designing and building D-RISC/Microgrids incrementally designed a processor architecture, based on an existing ISA and infrastructure?} Here the answer is only partially ``yes'': the design was indeed incremental (starting from a simple, known-to-work RISC pipeline) and used an existing ISA (Alpha), but it did not reuse an existing infrastructure. Instead, all the infrastructure for the project was built from scratch.

\paragraph{Did the process of designing and building D-RISC/Microgrids enabled the dynamic management and optimisation of available resources for a variety of goals such as performance, power and reliability?} The jury is still out on this one; no answer was given yet after many years of research. Even a published doctoral thesis on the topic~\cite{vantol.13}, which merely touched performance-driven resource management, did not yield definite answers. As of this writing, another doctoral candidate is working on the reliability issue, but issues of power are still left untouched.

\paragraph{Did the process of designing and building D-RISC/Microgrids include a formal definition of the architecture's execution properties?} Yes, namely in~\cite{tdvu.07.icfem} and~\cite[Chap.~7]{poss.12}.

\paragraph{Did the process of designing and building D-RISC/Microgrids include an incremental development of the architecture's infrastructure? (i.e. simulators, compilers, binary-to-binary translators and even silicon intellectual property)} Here the answer
is only partially ``yes.'' Simulators and compilers were developed,
and parts of silicon IP (cf.~\cite{danek.12}), however there were no
binary-to-binary translators produced to establish a compatibility
path with existing code, as initially envisioned.

\subsection{Restrospective on the research question}

Since the process of designing D-RISC/Microgrids did not exhibit all
the expected properties, it cannot be used to answer ``yes'' firmly to
that project's research question. However, ``no'' cannot be
confidently given either. In other words, the question is still mostly
left unanswered.

Instead, I can only summarize the situation by proposing that the NWO
simply funded some additional \emph{development} of D-RISC/Microgrids,
and the project's description only merely \emph{guided the development
  process} without pressuring it into delivering scientific output.

Moreover, it is clear to me that the original research question
\emph{was so ill-phrased that it cannot be answered scientifically}: I
cannot see any experimental path that would yield a definite answer
within a reasonable time frame. 

Consequently, any rephrasing would be equally improductive, for example
determine whether there is any \emph{other} set of concurrency
controls that yield a ``yes'' answer on the same research question, or
whether D-RISC/Microgrids can be fixed/enhanced to this aim. 

Therefore, in my opinion, the original phrasing for the project NWO
Microgrids cannot be used to motivate further work in this area.  The
corollary is that researchers should not exploit the past attention
given by NWO's to this question as justification to spend more effort
in this area. If justification is needed, it must be found somewhere
else.

\section{C. Jesshope, APC 2008}

This whitepaper/article~\cite{jesshope.08.apc} made a statement of
intent about the applicability and the aims of D-RISC/Microgrids. It
introduces the ``SVP model,'' an intellectual construction used from
2008 to 2011. SVP intended to abstract the specific inner workings of
D-RISC/Microgrids, keeping only the high-level semantics of its TMU
concurrency management protocol visible to programmers.

\begin{table}
\begin{tabular}{p{.48\textwidth}p{.48\textwidth}}
Problem & Proposed answer \\
\hline \hline
How to effectively program distributed multiprocessor systems & Use SVP's simple concurrency control primitives \\
\hline
How to make architectures that are both efficient and can tolerate a large latency in responding to external events &
Use a combination of native support for dataflow scheduling and split-phase transactions throughout the system, such as found
 in SVP implementations \\
\hline
How to design programming model that is both deterministic and free from deadlock under concurrent composition & Use SVP's strictly
hierarchical concurrency and forward-only communication patterns \\
\hline
How to ensure binary compatibility across a range of implementations from a single processor to the highest level of concurrency a particular application can support & Use SVP's granularity-independent abstraction of concurrency resources \\
\end{tabular}
\caption{Problems purportedly solved by SVP.}\label{tab:mgprob}
\end{table}

According to this article, D-RISC/Microgrids as abstracted in SVP
should have solved the problems listed in \cref{tab:mgprob}. 
Note that this article defined SVP and its benefits \emph{before D-RISC's TMU, and thus Microgrids were fully defined.} 
As it happened, the advertised features of SVP 
ended up \emph{not being implementable in D-RISC/Microgrids}. Specifically:

\begin{itemize}
\item end-to-end asynchrony stops at the chip boundary, both at
the memory and I/O interfaces, and these latencies cannot be fully
tolerated;
\item the lack of hardware mechanisms to virtualize resources prevented the
proper implementation of deadlock-free composition.
\end{itemize}

Moreover, although binary compatibility is possible across chip
technologies, the execution performance of the code ended up not being
portable between different number of cores and interconnect
topologies. 

Since 2011, when D-RISC's TMU was well-enough defined that it was
both obviously \emph{different from and more powerful than SVP's abstractions},
the SVP model has been downplayed and is not a central component of
publications any more.

\section{EU Apple-CORE: 2008-2011}

The project Apple-CORE was funded by the European Union based on a
statement of intent via an \emph{abstract}, and via a list of explicit 
\emph{objectives}. There was no ``research question'' per se, as
the goal of the project was to build infrastructure and show
that the objectives were reached as a consequence.

\subsection{Summary of outcomes}

To an outsider, 
the EU seemed to have funded research to develop a new general-purpose processor,
that would extend and possibly even replace the technology currently
in use in commodity hardware. Had that objective been reached, the
project would have been disruptive indeed. This \emph{potential} to
both disrupt the state of the art and advance technology in a way
largely beneficial to society was sufficient, to the proposal's
initial reviewers, to justify the investment.

However, there is an ethical issue at hand. First, the Apple-CORE
proposal stated that D-RISC and Microgrids were already
designed and a code generator for $\mu$TC was available prior
to the start of the project, whereas it was known to the authors of the proposal that
this was not true. 

Only after the first year, after reviewers had been induced to believe
the issues were minor, did it become clear that the EU was also
funding this prerequisite technology. Also, at that point there was no
evidence that this ``initial'' technology would be sufficient to
research all the project's original objectives in a timely fashion.
Indeed, what happened is that the overhead of producing these
prerequisites prevented the consortium from exploring all the issues.

In short, Apple-CORE \emph{did not actually have the potential to
  disrupt the state of the art and advance technology in the way
  announced using the budget requested}. Besides, the details of scientific
outcomes (\cf next section) do not reveal any strong evidence that the
Apple-CORE technology can replace existing processors (\cf also \cref{sec:incompat}). 

\paragraph{Disclaimer}
{\itshape The results and circumstances described below have been
brought to the attention of the project's reviewers while
the project was ongoing, and the project was judged successful
by both the reviewers and the project officer \emph{despite} these
issues. As I have learned since then, most
large, publicly-funded projects suffer from more serious issues and
the issues described here pale in comparison.}

\subsection{Outcomes \vs project objectives}

I present below how the project's outcome, as can be observed at the
end of 2012, relates to each stated objective in the project's
description of work. I list the objectives in decreasing order of
success, and mark with ``$\filledmedtriangleright\filledmedtriangleright\filledmedtriangleright$'' those
points that most diverge from the overall initial goal of the project.

\paragraph{Apple-CORE will investigate the support structures in 
implementing the SVP model in the LEON 3 processor and develop an SVP soft-core prototype.}
This was achieved, and was quite successful~\cite{danek.10.ddecs,sykora.11.lncs,danek.12}.

\paragraph{Apple-CORE will investigate the integration of instruction-set extensions to support custom accelerators based on both microthreads and families of microthreads.}
This was achieved, and was quite successful~\cite{danek.12}.

\paragraph{Apple-CORE will explore the gains of SVP in the context of data-parallel 
programming, investigate the implications of functional concurrency and explore the possible design space.}
This was achieved, and was quite successful~\cite{a-c-d44}.

\paragraph{Apple-CORE will support many-core processors by 
capturing concurrency systematically using 
instructions in the processors’ ISA and by dynamically mapping 
and scheduling that concurrency in the processors’ implementation (the SVP model).}
This was achieved~\cite{poss.12.dsd}.

\paragraph{Apple-CORE will derive a set of loop transformations to transform iterative computations into a combination of independent and dependent families of threads respecting the communication restrictions in the SVP model.}
This was achieved~\cite{saougkos.09.cpc,saougkos.11}.

\paragraph{Apple-CORE will extract task, loop and, implicitly in the SVP model, instruction level concurrency in the parallelising C compiler.}
This was only partially achieved: only loop and ILP was extracted. Task concurrency wasn't.

\paragraph{
$\filledmedtriangleright\filledmedtriangleright\filledmedtriangleright$
Apple-CORE will provide binary-code compatibility across generations of multi-cores from few- to many-cores.}
This was achieved, although Apple-CORE also showed that code that performs well on
a small number of cores typically does not scale (performance- and
efficiency-wise) to large number of cores. Conversely, code that runs
with interesting speedups on large number of cores do not run
efficiently on small number of cores. The binary compatibility is thus
merely functional: it is possible to run the same code and obtain the
same results, but the performance is not portable. One can easily
argue that this form of compatibility is not really what was desired.

\paragraph{
$\filledmedtriangleright\filledmedtriangleright\filledmedtriangleright$
Apple-CORE will implement and evaluate memory models and coherency protocols for many- core systems.}
This was achieved, only to conclude that the proposed models and
protocols were cumbersome to use, inefficient and otherwise
detrimental to performance for any configuration larger than 30-60
cores.

\paragraph{
$\filledmedtriangleright\filledmedtriangleright\filledmedtriangleright$
Apple-CORE will study the resource management issues that are exposed in exploiting massive concurrency as it arises from data-parallel or functional program specifications.}
This was achieved: the management issues were indeed studied, but only
to conclude that the Apple-CORE strategy did not significantly
simplify the problem, which is otherwise shared by all research projects
in this field.

\paragraph{
$\filledmedtriangleright\filledmedtriangleright\filledmedtriangleright$
Apple-CORE will provide high-level programming environments that
improve the programming productivity and automate the generation of
concurrency, or at least separate the concerns of concurrent
programming from its implementation, i.e. automate all scheduling and
synchronisation.}  This was only partly achieved. By funding extra
development on Single-Assignment C, Apple-CORE did indeed ``improve
the programming productivity and automate the generation of
concurrency.'' However this effort was not directly related to
D-RISC/Microgrids: the improvements on SaC are portable to any
parallel hardware supported by the SaC compiler. Furthermore
Apple-CORE did not ``separate the concerns of concurrent programming
from its implementation,'' and neither did it ``automate all
synchronization.''

\paragraph{
$\filledmedtriangleright\filledmedtriangleright\filledmedtriangleright$
Apple-CORE will investigate and implement memory protection and security issues for many- core systems.}
This was investigated but not implemented.

\paragraph{
$\filledmedtriangleright\filledmedtriangleright\filledmedtriangleright$
Apple-CORE will implement a port of a micro-kernel operating system onto one or more of the processor platforms (emulation and/or soft core).}
This was not investigated and not achieved, \cf also \cref{sec:incompat}.

\paragraph{
Apple-CORE will promote the $\mu$TC language as a standard front-end
to the gcc compiler and will use it as a target for all user-level
compiler development.}  This did not occur, and $\mu$TC is not being
used any more, for the reasons presented in~\cite[App.~G]{poss.12}. Instead,
the project used another front-end to D-RISC/Microgrids called SL~\cite{poss.12.sl}, which
is riddled with practical limitations and has yet to gain credentials in the scientific community.

\paragraph{
$\filledmedtriangleright\filledmedtriangleright\filledmedtriangleright$
Apple-CORE will investigate and evaluate programming productivity issues for the tools developed.}
This was not investigated nor evaluated.

\paragraph{
$\filledmedtriangleright\filledmedtriangleright\filledmedtriangleright$
Apple-CORE will select a range of benchmark applications of interest to potential users of the SVP model within the European computer industry.}
This was only partly achieved: the industrial participation in the project was low, therefore the relevance of the resulting benchmark selection
cannot be confidently ascertained.

\paragraph{Apple-CORE will build an infrastructure of tools that will enable the SVP model to be evaluated and adopted by the European Computer Industry.}
This objective is untestable. Although an infrastructure of tools was produced, no adoption by the European Computer Industry has yet occurred.

\subsection{Outcomes \vs intents in the project's abstract}

The project's abstract also declared research intents not
covered otherwise in the objectives. I review them here:

\paragraph{The benefits are large, [...] as compilers need only capture
concurrency in a virtual way rather than capturing, mapping and
scheduling it.}
These benefits were not observed. Instead, Apple-CORE taught us is not sufficient to capture concurrency;
some semantics that brings an intuition of the machine back to the programmer
was necessary after all.

\paragraph{This separates the
concerns of programming and concurrency engineering and opens the door
for successful parallelising compilers.}  There was no breakthrough in
programming and concurrency engineering during Apple-CORE, and no
``successful parallelising compiler'' has been produced as a
result. Technically, there were parallelising compilers produced, but
they are not yet ``successful'' insofar they have not yet gathered any
user base other than their own developers.

\paragraph{Particular benefits can be expected for data-parallel
and functional programming languages as they expose their concurrency
in a way that can be easily captured by a compiler.}
This was indeed shown, although this can be equally shown using
most parallel platforms in this day and age.

\paragraph{Another advantage
of this approach is the binary compatibility the new processor has
with the modified ISA. [...]  Once code is compiled with the new
tools, binary-code is executable on an arbitrary numbers of processors
and hence provides future binary-code compatibility.} See above: although
the code is binary-compatible, the performance is not portable.

\paragraph{The concurrency controls also allow for management of
partial failure, which together with the binary-code compatibility
provide the necessary support for reliable systems.}  This was
not shown in practice by Apple-CORE, although Apple-CORE's
infrastructure does simplify a research project on this topic,
started later on.

\paragraph{
$\filledmedtriangleright\filledmedtriangleright\filledmedtriangleright$
Finally, this
approach exposes information about the work to be executed on each
processor and how much can be executed at any given time.
This
information can provide powerful mechanisms for the management of
power by load balancing processors based on clock/frequency
scaling. }
This was not researched in Apple-CORE.

\paragraph{
$\filledmedtriangleright\filledmedtriangleright\filledmedtriangleright$
  In particular, the binary compatibility provides a unique
  opportunity to make an impact on commodity processors in Europe.}
This was not achieved, as there is no binary compatibility with
existing processors. Actually, the argument of ``binary
compatibility'' throughout the project proposal diverges from usual
expectations. ``Binary compatibility'' is usually understood to mean
``backward compatibility with existing binary code,'' meaning that
existing software from other platforms can be reused on the new
platform. However, Apple-CORE instead promotes ``binary compatibility
between multiple instances of the Apple-CORE technology,'' \ie no
binary backward compatibility with other platforms.

\section{C. Jesshope et al., ParCo 2009}

This article~\cite{jesshope.09.parco} rephrased the motivation
behind the D-RISC/Microgrid work, two years in the Apple-CORE project:

\begin{quote}
\itshape In a more general market [than embedded and special-purpose
  accelerators], the labour-intensive approach of hand mapping an
application is not feasible, as the effort required is large and
compounded by the many different applications. {\bfseries A more automated
approach from the tool chain is necessary. This investment in the tool
chain, in turn, demands an abstract target to avoid these
compatibility issues. That target or concurrency model then needs to
be implemented on a variety of platforms to give portability, whatever
the granularity of that platform.  Our experience suggests that an
abstract target should adopt concurrent rather than sequential
composition, but admit a well-defined sequential schedule. It must
capture locality without specifying explicit communication. Ideally,
it should support asynchrony using data-driven scheduling to allow for
high latency operations. However, above all, it must provide safe
program composition, i.e. guaranteed freedom from deadlock when two
concurrent programs are combined. Our SVP model is designed to meet
all of these requirements.} Whether it is implemented in the ISA of a
conventional core, as described here or encapsulated as a software API
will only effect the parameters described above, which in turn will
determine at what level of granularity one moves from parallel to
sequential execution of the same code.
\end{quote}

A few years afterwards, is the D-RISC/Microgrids management protocol
matching the claims? I review them here in decreasing order of
success.

\paragraph{The model should support asynchrony using data-driven scheduling to
allow for high-latency operations.} This was achieved (primary feature of D-RISC).

\paragraph{The model should adopt concurrent rather than sequential composition.}
This was achieved, although \emph{both} concurrent and sequential composition are equally promoted.

\paragraph{The model must admit a well-defined sequential schedule.}
This was mostly achieved~\cite[Chap.~10]{poss.12}. A sequential schedule is not properly defined
as soon as a program manipulates on-chip resources (in particular cores) explicitly.

\paragraph{The model needs to be implemented on a variety of platforms.} This was not achieved. 
A software emulation of D-RISC's \emph{envisioned} TMU was implemented early on~\cite{tol.09.jsa}, but
the actual D-RISC TMU ended up with different semantics which have not yet been implemented elsewhere.

\paragraph{The model must capture locality without specifying explicit communication.}
This was not achieved: explicit communication is required between different threads.

\paragraph{Above all, it must provide safe program composition, i.e. guaranteed freedom from deadlock
when two concurrent programs are combined.}
This was not achieved: the ability of code to manipulate resources explicitly, combined with the lack
of full resource virtualization, may cause compositions to deadlock from resource starvation. 

\section{ASCI 5-year research plan, 2010}

This document was submitted to a consortium of Dutch universities
at the start of 2010, to define the overall research plan of the consortium
over the period 2010-2014.

Over D-RISC/Microgrids this report states:

\begin{quote}
\itshape
Our work on fine-grain threaded architectures with data-driven
scheduling using the SVP concurrency model will continue but we will
also explore software implementations of SVP on other emerging
multi-core architectures such as Niagara and Intel's SCC. This will
allow us to explore multi-grain architecture and develop an
infrastructure to support such an approach. One of the major
directions in this work will be the development of a coherent set of
operating system services that support space sharing in these
heterogeneous environments and yet provide a secure operating
environment that can be scaled from chip-level micro-grids to globally
distributed Grids.  One of the major challenges, especially in
mainstream computing, will be in making these systems programmable
without specialized concurrency knowledge and we have designed
programming language support to express parallel computations and
systems at a very high level of abstraction and developed compilation
technologies that effectively map the abstract descriptions to
concurrent computing environments. We have international
collaborations developing the functional, data parallel language SAC
(Single Assignment C) and the asynchronous co-ordination language
S-Net. Our long-term vision is in the direction of a compilation
infrastructure that automatically adapts running programs derived from
high-level specifications to a heterogeneous and dynamically varying
execution environment based on continuous reflection of execution
parameters.
\end{quote}

To this date, SVP was not implemented on other architectures. No
operating systems services have yet developed that support space
sharing in heterogeneous environments and provide a secure exeuction
environment. The D-RISC/Microgrids language tools are not yet able to
map the abstraction of concurrent resources to maximize performance
and efficiency. ``Automatic adaption of running programs towards
heterogeneous and dynamically varying execution parameters based
on continuous reflection'' was not achieved either yet.

\begin{summary}
\begin{itemize}
\item 
The D-RISC/Microgrids project was purportedly intended to solve major issues in
micro-architecture research, related to scalability in performance and
efficiency in general-purpose microprocessors. 
\item
The strategy to solve these issues was to implement a
combination of dataflow scheduling with hardware support for thread
concurrency management within and across cores on chip. 
\item
Implementation
was carried out, but the results are inconclusive. On the one hand, the proposed
hardware does indeed provide higher performance and efficiency in
regular, data-parallel workloads, but these are also the ``boring''
applications which benefit equally well from vector units or accelerators
in conventional processors. On the other hand, no evidence has yet
been produced that the proposed hardware benefits larger applications
with more irregular workloads. 
\item
Power efficiency and intelligent resource
management was regularly advertised but not actively researched.
\item
Effort has been invested into widening the scope of the technology
towards applications and industrial relevance, but these applications
have not yet materialized.
\end{itemize}
\end{summary}

\chapter{Methodology issues}

What the body of published and unpublished materials reveal is a
large, loosely scoped enterprise to define a multi-core processor chip
with the ambitious aim to solve the most significant problems of
architecture research in the period 2010-2020.

\section{Actual methodology}

Although never explicit, a strategy guides the effort:

\begin{enumerate}
\item accumulate technology around the simple ideas of \emph{dataflow
  scheduling} and \emph{partial hardware support for concurrency
  management}, so as to define an execution platform able to run
  parallel benchmark programs;
\item ``try it out'' and measure how it behaves;
\item if the measurements are unsatisfactory, return to step \#1;
  otherwise publish results and claim success.
\end{enumerate}

\begin{figure}
\centering
\includegraphics[scale=.4]{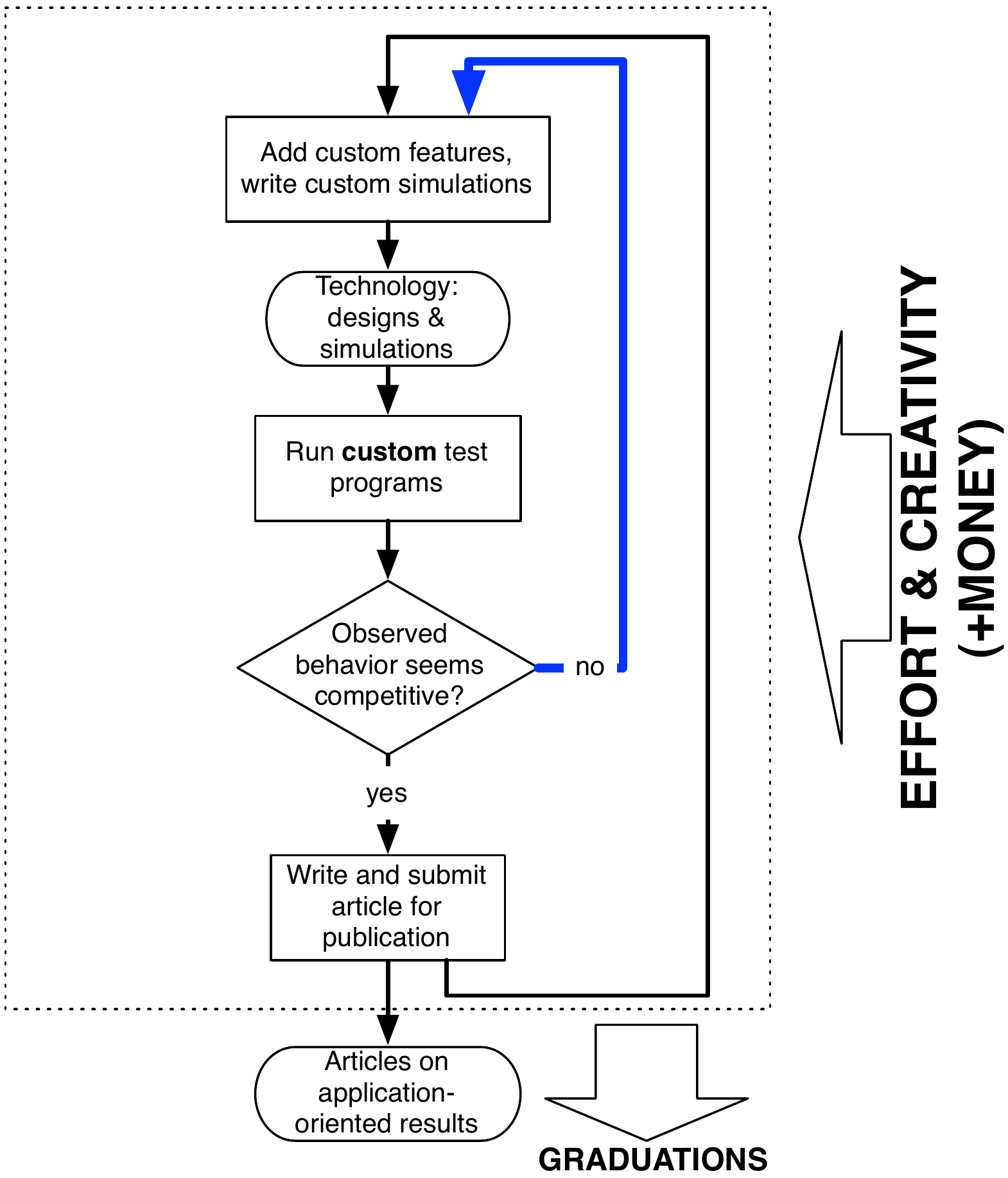}
\caption{High-level overview of research dynamics around D-RISC/Microgrids.}\label{fig:mgproc}
\end{figure}

An overview of this process is given in \cref{fig:mgproc}. From an
outsider's perspective, this research activity is independent, as its
only input is human effort and financial investment. Its overall
output is scientific articles on measured results, and a regular stream
of educated practitioners.

Over the year, two patterns have emerged. The first is that the
research group often stalls, busy looping from unsatisfactory results
back to implementing more features without questioning the overall
strategy (thick blue arrow in the figure). The consequence is an irregular, unfocused publication
throughput and doctoral candidates abandoning their research from lack
of focus. The second pattern is that only positive, ``competitive''
results are retained as candidates for publication. The consequence is
a lack of visibility on the research process, methodologies and
shortcomings, although these could also be useful and valuable to the
scientific community.

\begin{figure}
\centering
\includegraphics[width=\textwidth]{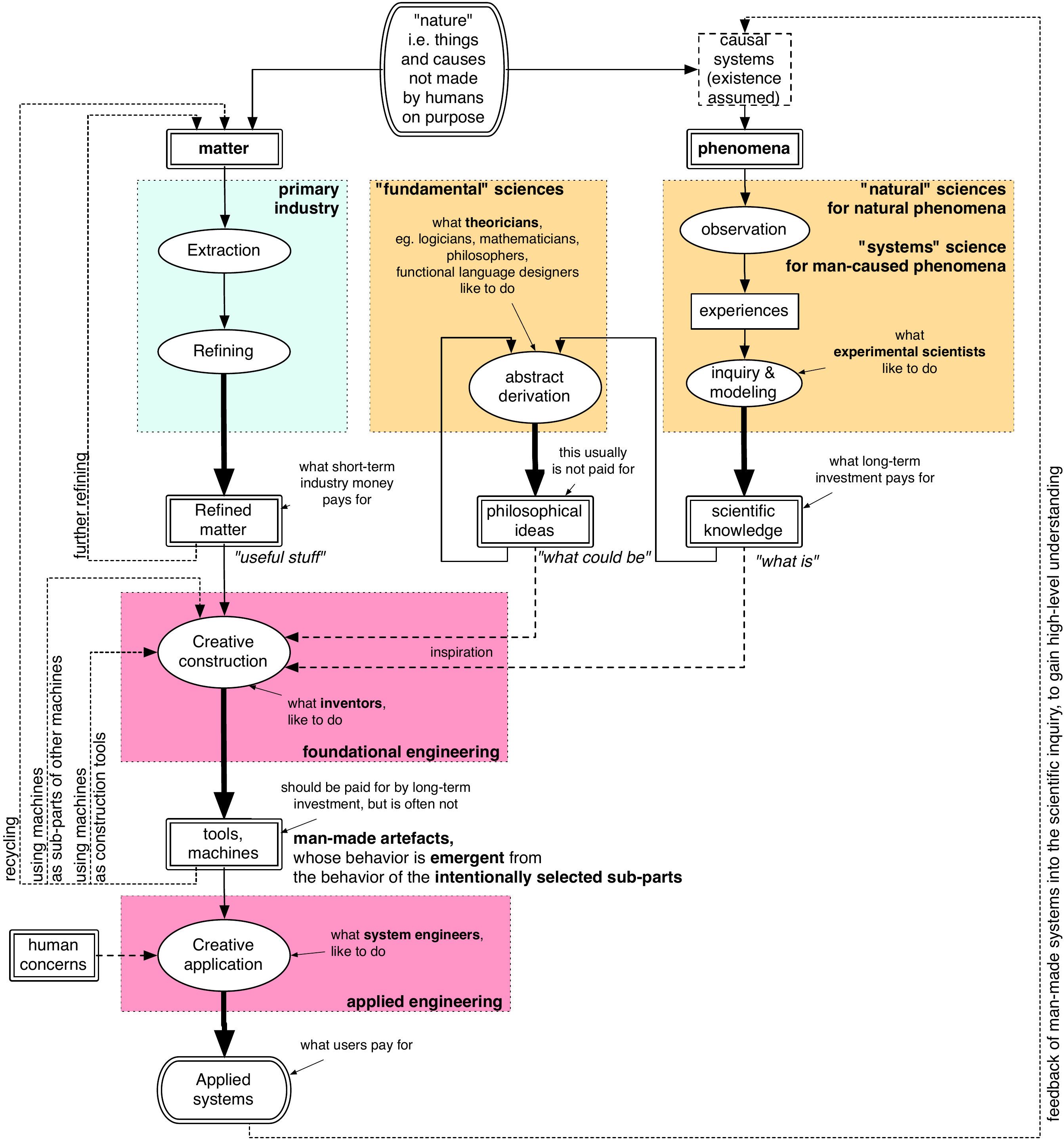}
\caption{Activities related to science.}\label{fig:science}
\end{figure}

\section{Relationship with other scientific activities}

Compare the process above with \cref{fig:science}, explained as follows in \cite[Chap.~1]{poss.12}:

\begin{quote}
\itshape
The traditional purpose of the fundamental sciences is the acquisition
of new knowledge pertaining to observed phenomena, in an 
attempt to describe ``what is.'' In parallel to the discovery of
new knowledge through scientific inquiry, philosophers, or
theoreticians, derive ideas of ``what could be.'' Via
formalisms, they construct structures of thought to validate
these ideas and derive iteratively new ideas from them.

We can focus for a moment on the human dynamics around these
activities. On the one hand, the intellectual pleasure that
internally motivates the human scientists is mostly to be found
in the acquisition of knowledge and ideas. For natural scientists, the
focus is on accuracy relative to the observed phenomena, whereas for
philosophers the focus is on consistency. On the other hand, the
external motivation for all fields of science, which materially
sustains their activities, is the need of humans for either
discovery or material benefits to their physical
existence.  From this position, the outcome of
scientific inquiry and philosophical thought, namely knowledge and
ideas, is not directly what human audiences are interested in.  The
``missing link'' between scientific insight and its practical benefits
is \emph{innovation}, an  \emph{engineering} process in two steps.

The first step of innovation is \emph{foundational engineering}: the
creative, nearly artistic process where humans find a new way to
assemble parts into a more complex artifact, following the inspiration
and foreshadowing of their past study of knowledge and ideas, and
guided by human-centered concerns.  Foundational engineering, as
an activity, consumes refined matter from the
physical world and produces new more complex things, usually tools and
machines, whose function and behavior are intricate, emergent
composition of their parts. The novelty factor is key: the outcome
must have characteristics yet unseen to qualify as foundational;
merely reproducing the object would just qualify as manufacturing.
The characteristic human factor in this foundational step
is \emph{creativity}, which corresponds to the serendipitously
successful, mostly irrationally motivated selection of ideas,
knowledge and material components in a way that only reveals itself as
useful, and thus can only be justified, \textsl{a posteriori}.

The other step is \emph{applicative engineering}, where humans
assemble artifacts previously engineerd into complex systems that
satisfy the needs of fellow humans. In contrast to foundational
engineering, the characteristic human factor here is meticulousness in
the realization and scrupulousness in recognizing and following an
audience's expectations---if not fabricating them on the spot.

The entire system of activities around science is driven by
a \emph{demand for applications}: the need of mankind to improve its
condition creates demand for man-made systems that solve its problems,
which in turn creates demand for new sorts of devices and artifacts to
construct these systems, which in turn creates demand for basic
materials as input, on the one hand, and intellectual diversity and
background in the form of knowledge and ideas. We illustrate this general view 
in \cref{fig:science} [...].  The role of education, in turn, is
to act as a glue, ensuring that the output of the various activities
are duly and faithfully communicated to the interested parties.
\end{quote}

In this context, the activity around D-RISC/Microgrids can be
recognized to actually constitute \emph{foundational engineering}: the process
of invention that produces tools and artefacts that can subsequently
solve ``real-world'' problems. 

This immediately highlights two major issues:
\begin{itemize}
\item the output of foundational engineering is measured by the tools
  and artefacts it produces, not merely their description in the form
  of academic publications. For the effort on D-RISC/Microgrids to be
  recognized and valued as innovation, it must be accompanied by the
  marketing of its \emph{technology}, including its flaws and
  limitations, and ultimately exploitation to real-world applications.
\item it is not the primary purpose of the academic institutions of
  science to fund and support foundational engineering.  Although it
  is not uncommon to see foundational engineering occur in academic
  environments, it is usually only accepted as a by-product
  exploitation of the other activities of science, namely natural and
  fundamental sciences. To justify continued effort on D-RISC/Microgrids 
  in a university research group, the fundamental principles of the technology
  must be extracted, abstracted, studied formally and generalized
  in a relevant, technology-independent fashion. This work has not been performed so far
  other than via the three isolated publications \cite{tdvu.07.icfem,vu.08.icamst} and \cite[Chap.~7]{poss.12}.
\end{itemize}

\begin{summary}
\begin{itemize}
\item Research on D-RISC/Microgrids is not following the scientific method.
It is instead currently carried out as an engineering enterprise, but without clear
technology ouputs and without identifying
its potential applications.
\item
Its relevance in a university research group is thus questionable.
\end{itemize}
\end{summary}

\chapter{Obstacles to further progress}

In \cref{chap:outcomes} I have presented the status of the research so
far, and identified the areas where it did not deliver on its own
self-set expectations. In this chapter, I follow up by reviewing the
likely causes of these shortcomings.

\section{Large project scope}

The project's aim is to eventually deliver a full general-purpose processor. However, the design
of a new processor architecture requires large research and implementation investment on \emph{all} the following fronts:

\begin{enumerate}
\item micro-processor logic (per core); \label{e:mp}
\item NoC interconnect: flow control, routing, failure management; \label{e:net}
\item cache management and inter-cache coherency; \label{e:caches}
\item core-NoC interfaces;  \label{e:if}
\item external memory interfaces; 
\item external I/O interfaces;
\item hardware/software interface design: ISA, but also I/O address space layout, MMU control, access to performance counters, etc; \label{e:isa}
\item ISA code generation in low-level compilers; \label{e:cgen}
\item architecture-specific optimizations in compilers; \label{e:opt}
\item architecture-specific support in software operating systems and programming language libraries; \label{e:os}
\item individual component validation, both from formal analysis and unit testing; \label{e:val}
\item component-level, then system-level modeling and simulation; \label{e:sim}
\item design of the integration strategy in a larger system (exploring system parameters and how system-level interconnects will impact intra-chip behavior);
\item circuit synthesis and prototype production (either ASIC or FPGA); \label{e:impl}
\item show casing and marketing via extensive comparisons with competition products; \label{e:cmp}
\item seeking and partnering with industry to define marketable products. \label{e:ind}
\end{enumerate}

Each of these areas would require multiple man-years worth of
investment before evidence of success in reaching the aim ``delivering a
full general-purpose processor'' can be strongly claimed.

To this date, the research effort around D-RISC/Microgrids was focused
on \cref{e:mp,e:caches,e:sim,e:isa}, although for \cref{e:isa} no clear
picture of the interface to virtual memory and I/O has yet emerged.  A
lot of effort was also spent towards \cref{e:cgen,e:os} to enable
benchmarking, albeit half-heartedly because the research group lacked
expertise in these areas until recently. Some effort was spent on
\cref{e:impl} with a research partner~\cite{danek.12}, although the resulting FPGA model only
implements 1 core connected to a bus, \ie it only shows limited
benefits of latency tolerance and does not implement the TMU's
multi-core coordination features. The other items have not been
investigated yet.  Even if the scope of the research was reduced to
``design a Microgrid component that can be integrated as accelerator
in a larger processor chip,'' as advertised in \cite{poss.12.dsd},
effort should still be invested on
\cref{e:net,e:if,e:opt,e:val,e:ind,e:cmp}. This is out of the reach of
an isolated university research group expected to deliver strong
results in periods of five years with a bandwidth of 1 to 5
contributors per year.

\section{``Here Is My Chip, You Figure It Out''}

Following the original aim thus requires partnerships with other
organizations to carry out the work together. However, in the
scientific community, partnerships only come into existence based on
mutually beneficial arrangements: a peer researcher or instution may
be willing to contribute effort and technology towards the betterment
of D-RISC/Microgrids, only if they get something in return. 

However, most of the interactions with potential partners so far were
carried out thus: ``here is our technology, we think it is good for
such-and-such use cases, what about you try to use it and tell us what
you do with it?'' For the reasons discussed extensively
in~\cite[Sect.~1.5, 15.1 \& Chap.~16]{poss.12}, this approach is
subject to the ``Here Is My Chip, You Figure It Out'' (HIMCYFIO)
hazard: faced with alien, unrecognizable technology, a potential
partner or user will be reluctant to invest the effort necessary to
cross the \emph{comprehension} threshold, even before they start to
think about potential shared endeavours. The producer of the
technology must cross this threshold preemptively to avoid the
HIMCYFIO pitfall.

Although my thesis \cite{poss.12} made one step in that direction,
further work is needed by this research group to bring the
D-RISC/Microgrid technology ``to the level'' of its potential
partners.  The participants must identify what challenges potential
partners are facing, and preemptively shape the technology into a
palatable solution to the partners' problems. 

To this day, the issues faced by the scientific peers in the same
research domain have not been investigated thoroughly by the research
group. The main reason for this is not lack of intent; rather a
crucial technical practical/historical obstacle: validation and peer
recognition in the micro-architecture community heavily relies on the
ability to exchange hardware platforms \emph{without modifying the
  software}, so as to enable sound comparisons between
solutions. Although source-level compatibility is sufficient
(recompiling code towards a new platform has become acceptable in the
community), it is also necessary: given the large effort necessary to
design and deliver hardware, little effort can be spent
rewriting/adapting benchmark code, which has often accumulated
man-years of design, towards new platforms. However, for the reasons
described below and as of this writing, D-RISC/Microgrid's
\emph{cannot be made source-compatible} with most existing benchmarks.
This limitation is a high obstacle to publication and thus visibility,
and cuts short most opening discussions with potential partners.

\section{Lack of features needed for compatibility}\label{sec:incompat}

In my opinion, the main obstacle to the usability of D-RISC/Microgrids
by third parties is the lack of the following features:

\begin{itemize}\bfseries
\item \emph{process virtualization}, including per-process virtual memory address spaces and virtual I/O channels;
\item interrupt-like \emph{mechanisms to handle faults and unexpected external events};
\item the \emph{ability to stop a process and inspect it} externally, \eg using a debugger;
\item the ability to \emph{preempt a running program and reclaim its resources}.
\end{itemize}

Programmers, in particular operating system and language implementers,
have been accustomed in the last 50 years to expect these features from any
general-purpose computer; the corollary is that all the existing
\emph{operating software} underlying existing application software
makes pervasive use of these services from the hardware platform.

The standpoint of the research group was that these features
are an ``historical artefact'' that were motivated at a time when
on-chip resources (cores, memory) were limited, and must thus be reconsidered at an
age where there are thousands of cores on chip and 64-bit addresses to
memory. As suggested in~\cite{tol.06,jesshope.08.samos,tol.11.apc,vantol.13}
and \cite[Sect.~3.3.2 \& Chap.~14--15]{poss.12}, this research group's public answer to
queries about these features goes as follows:
\begin{itemize}
\item ``preemptive event handlers should not be needed when any two
  system-level tasks waiting on events could be active at the same
  time in different hardware threads or cores (of which there are
  thousands available);''
\item ``separate virtual address spaces should not be needed when a single virtual 64-bit
  address space can be partitioned a thousand-fold while still
  providing petabytes addressable to each process;''
\item ``once processes are allocated over space and not over time,
  process boundaries are congruent to areas on chip and resource reclamation can be implemented simply by fully resetting the
  corresponding hardware resources;''
\item ``issues of debugging and investigation do not need special supports as long
  as simulators and emulators are available: debugging can be performed
  from the simulator/emulator's host.''
\end{itemize}

As an insider, I can also report the second half of the answer: it is
possible to add support for these features, but the fear is that doing
so would make the research more difficult because a larger set of
issues would need to be considered. An ungrounded assumption is that
introducing support for virtualization would introduce overheads in
the hardware and make D-RISC and Microgrids less competitive against
other processors, including for the workloads where it currently
shines.  The assumption is ungrounded because the consequences of
extending D-RISC in that direction have simply not been investigated
yet.

{\bfseries
In short, the position of the research group is ``these are
complicated engineering issues, but we think they are only superficial
usability concerns so they do not deserve our attention yet.''
}

The net effect is that existing operating software cannot be reused
with the proposed platform. Even assuming that new, custom-built
operating software \emph{could potentially show} that these
traditional mechanisms for virtualization can be avoided, the very
lack of software compatibility caused by the current situation may
well form the unsurmountable barrier preventing the group from forming
the partnerships needed for further developments, for the reasons
outlined in the previous section.

(For the sake of clarity, according to my own analysis these features
are both necessary and sufficient to immediately enable porting and
reusing operating software and existing major programming framework on
the proposed platform. To my knowledge, these features constitute
exactly the remaining obstacle to compatibility.)

\section{Weakly grounded and tested hypotheses}

The foundation for the ``D-RISC enterprise'' is the observation
that the static and dynamic costs of OoOE in GP processors is largely caused
by the logic necessary to discover instruction-level parallelism at run-time.
The hypotheses of the D-RISC/Microgrids research are then articulated as follows:
\begin{enumerate}\itshape
\item By shifting the responsibility to
discover concurrency, from the run-time to design-time (or compile-time),
these costs can be avoided. And instead of encoding concurrency the
VLIW way, which is weak when faced with the unpredictable variance of on-chip access
latencies in large chips, the concurrency can be encoded via threads
instead.
\item If thread creation is not more expensive than simple branches,
  many sequential patterns including function calls and loops can be
  transparently replaced by threads, and this
  transformation can be performed casually in code generators for any
  language. As a result, just using the platform's compilation tools 
  can introduce concurrency automatically in any sequential software and
  solve the general ``programmability'' challenge of parallel hardware.
\item If concurrency management is encoded in the ISA with lightweight
  instructions, the same binary code can be run under any amount of
  resources, starting with a single thread on 1 core where it can be
  as fast as an equivalent branch-based sequential code.
\item If concurrency management is encoded without explicit reference to parallel
 hardware resources (``resource-agnostic''), the execution platform
 can adapt the code at run-time to maximize performance and efficiency 
 to the resources effectively available.
\end{enumerate}

The first hypothesis has been largely confirmed to hold for large
classes of applications, but then not by D-RISC specifically: Intel's
HyperThreaded cores, then Sun/Oracle's Niagara cores, have been endorsing
the benefits of hardware multithreading for general-purpose computing
(especially in the datacenter domain) for a few years already.

The other hypotheses are more problematic. 

The 2nd hypothesis heavily relies on the assumption that the
sequential composition of activation frames (for function calls and
loop iterations) can be cheaply and automatically replaced by parallel
composition. As I reveal partially in~\cite[Chap.~9]{poss.12}, this
research group had ``forgotten'' to consider that
\emph{general-purpose computations may use arbitrarily large amounts
  of storage} at any step. This implies a local, ``private'' memory
area \emph{of varying size} for each activation frame \emph{while it
  is running}.

With sequential execution, this is possible to implement cheaply with
a stack because only the most recently called activation frame is
running and can use the entire remaining memory. With parallel
execution, multiple activation frames are running and compete for the
available memory. Without much care (and thus ``intelligence'' that
must in turn be implemented in the TMU, raising its cost), the management of all these ``memory clients''
that all grow and shrink their local memory dynamically becomes a
bottleneck to parallel scalability.

Note that this problem is avoided in most highly-parallel
``accelerators'' by simply stating they do not support general-purpose
computations~\cite[Sect.~1.4\&12.6]{poss.12}, and in supercomputers by
providing large amounts of local RAM next to each processor. The
``supercomputer approach,'' applied to D-RISC/Microgrids, would imply
embedding large SRAM modules next to each core on chip or DRAM using
3D stacking, a direction not yet envisioned by this research group.

The 3rd hypothesis about automatic sequentialization was partly shown to hold with
D-RISC for threads that mostly perform local computations on
non-shared data and resources. As soon as shared or non-local
resources are used, sequentialization must choose a computation order
that maximizes locality of access and reuse. In sequential code,
typically the specified order of operations carries domain knowledge
from the programmer or compiler about locality; with threaded code,
this knowledge has disappeared from the code and must be reintroduced
at run-time while sequentializing. So far, the results produced show
that the automatic sequentialization is doing a poor
job~\cite[Sect.~10.4]{poss.12} but no research is being performed to
improve this state of affairs.

The 4th hypothesis about performance portability has been only
confirmed with D-RISC for data-parallel code with regular
data access patterns. For other types of concurrent code,
experimentation has shown that performance and efficiency are largely
dependent on the congruence between the data access patterns and
the physical topology of the interconnect between cores and
memory~\cite[Chap.~13]{poss.12}. However, to this date the concurrency
management interface of D-RISC's TMU does not allow to specify data
access patterns, so the TMU cannot make ``intelligent'' decisions
about placement. No further research is being performed in this
direction either.

All in all, these hypotheses seem \emph{attractive}: our peers have
reviewed these hypotheses through our publications and confirmed
implicitly, by accepting publication and funding further research,
that the hypotheses have merit and that our research efforts to test
them are scientifically worthwhile. However, I can also practically
observe that the work is not organized around strategic experiments
that would provide clear answers on these hypotheses. Meanwhile, I
have observed negative results that tend to invalidate the
hypotheses as they now stand, and I have also observed that these negative
results are not publicly exposed; instead, they are casually treated
as ``bugs'' and addressed by ad-hoc workarounds in the architecture
design.

\section{Competition and lack of differentiation}

Research on D-RISC/Microgrids was initiated in the late 1990's: a time
where multi-core chips were not yet widely used, and were there was
still a lot of uncertainty about concurrent programming. Back then,
D-RISC's approach was not only fresh, it was also spearheading
research in this area. There was thus not much to consider in terms of
``competition'' and ``related work.''

This has now changed. Since 2005, we have observed an explosion of
hardware multithreaded cores, multi-core chips and concurrent
programming frameworks. There now even exists architectures where
concurrency management is partly implemented in hardware: NVidia's
Fermi, Kalray's MPPA, Tilera's TILE are examples. Parallel and
multi-core programming is now being studied by students as a basic
course, and a wealth of software frameworks have evolved to manage
large number of ``micro threads'' efficiently on today's multi-core
chips: qthreads, codelets, green threads, Erlang's run-time system,
etc. Moreover, many high-level constructs to expose concurrency in
programming languages have been designed, e.g. in C/C++ (in the new
2011 standards), Scala, Haskell, etc. These have gradually introduced
\emph{expectations} in programmers' sub-cultures about what features
programming environments should and should not provide. Finally, some
architectural features have become widely accepted as fundamental to
the continued relevance of multi-cores, e.g.  transactional memory and
heterogeneity, which have yet not been analyzed nor picked up by the
D-RISC/Microgrids enterprise.

In short, the research field has become crowded. To attract attention
and thus gather momentum, it is essential to \emph{acknowledge the
  competition}, \emph{stay competitive} by keeping up and integrating
the good ideas from other projects, and simultaneously
\emph{differentiate} the new technology by pitting it against its
competition systematically. To this date, the research effort around
D-RISC/Microgrids has not focused on studying and integrating the
growing state-of-the-art, and differentiation is not expressed in
publications.

\begin{summary}
\begin{itemize}
\item 
The D-RISC/Microgrids project has the ambitious aim to produce a
general-purpose processor chip able to disrupt the current
state-of-the-art. However, the limited human resources dedicated to
the project are insufficient to reach this aim in isolation.
\item 
The expansion of the research group to a community of users and and research
partners is blocked by a fundamental lack of compatibility with existing operating
systems and application software.
\item This lack of
compatibility is not properly justified, neither by practical nor
theoretical reasons.
\item Meanwhile, the scientific effort to test the hypotheses that underly
the D-RISC/Microgrids project is poorly directed, and not
enough attention has been given to negative results that
invalidate these hypotheses.
\item Finally, the multi-core research field is nowadays much
more crowded than it was ten years ago, yet the research on D-RISC/Microgrids
does not acknowledge its competition nor attempts to differentiate
its contributions from the state of the art.
\end{itemize}
\end{summary}

\part{Reconstruction}

\chapter{Actual contributions}

\section{Concrete scientific contributions}

The scientific output of the project is positive on at least four angles.

First, some published discussion-oriented articles have
articulated interesting challenges to the tacit assumptions of the
architecture community about the exploitation of concurrency in processors,
\eg~\cite{bell.06.jpp,jesshope.08.samos,tol.11.apc,jesshope.09.parco,bernard.10.ppl,bernard.10,poss.10.amp,poss.12,vantol.13}.

Second, all results-oriented articles accepted by peer review
for publication are based on real experimental results using
``honest and best effort'' implementations of the proposed ideas,
\eg~\cite{luo.02,jesshope.09.arcs,hasasneh.07.jsa,vu.08.icamst,bousias.06.cj,bousias.09.jsa,poss.12.dsd}. 
Regardless of the conclusions drawn from them, these results
constitute a sound database of prior work to all future
researchers working on related areas.

Third, the research group has indirectly contributed to other projects
via its few partnerships. For example, the close work relationship
with the designers of Single-Assignment C and S-NET have yielded both
joint scientific
outputs~\cite{jesshope.08.apcsac,a-c-d44,poss.12.interact,grelck.09.cpc}
and technology improvements, directly or indirectly inspired by the
work on D-RISC/Microgrids.

Fourth, it the project has enabled ancillary research in novel techniques for system
simulation when cores are hardware multi-threaded~\cite{tol.09.jsa,tol.11.dutc,mirfan.11,poss.12.rapido,mirfan.12,lankamp.13.mgsim}, whose results are
scientific contributions on their own regardless of the specific merits of D-RISC/Microgrids.

\section{Technology products}

The research efforts have produced the following components and tools:
\begin{itemize}
\item svp-ptl and d-utc~\cite{tol.09.jsa,tol.11.dutc}, a library of TMU-like services implemented in software over POSIX threads,
  ready to implement concurrent software over multi-cores and distributed memory systems;
\item MGSim~\cite{poss.12.rapido,lankamp.13.mgsim}, a combination of:
\begin{itemize}
\item a general discrete-event, component-based simulation framework in C++, and
\item a library of component models that can be used to simulated D-RISC/Microgrid-based architectures;
\end{itemize}
\item HLSim~\cite{mirfan.11,mirfan.12}, a discrete-event, thread-based simulation of multi-scale systems using the TMU protocol
  and the API from svp-ptl;
\item the ``SL core'' package~\cite{poss.12.sl}, a combination of:
\begin{itemize}
\item a code translator from SL to D-RISC code, using any of its 3 possible ISAs,
\item a code translator from SL to the API of HLSim and svp-ptl,
\item a code translator from SL to ``vanilla'', sequential ISO C,
\item an incomplete port of a standard C library suitable for use in the simulated D-RISC/Microgrid environments, and
\item operating system components for resource management and interfacing with I/O services on D-RISC cores;
\end{itemize}
\item a set of micro-benchmarks using the SL language extensions that exercise the architecture and demonstrate the features
  of the simulation frameworks;
\item via research partners, the UTLEON3 core design in VHDL~\cite{danek.12} which implements one D-RISC core with a partial TMU
 for use on FPGA.
\end{itemize}

\begin{figure}
\centering
\includegraphics[scale=.6]{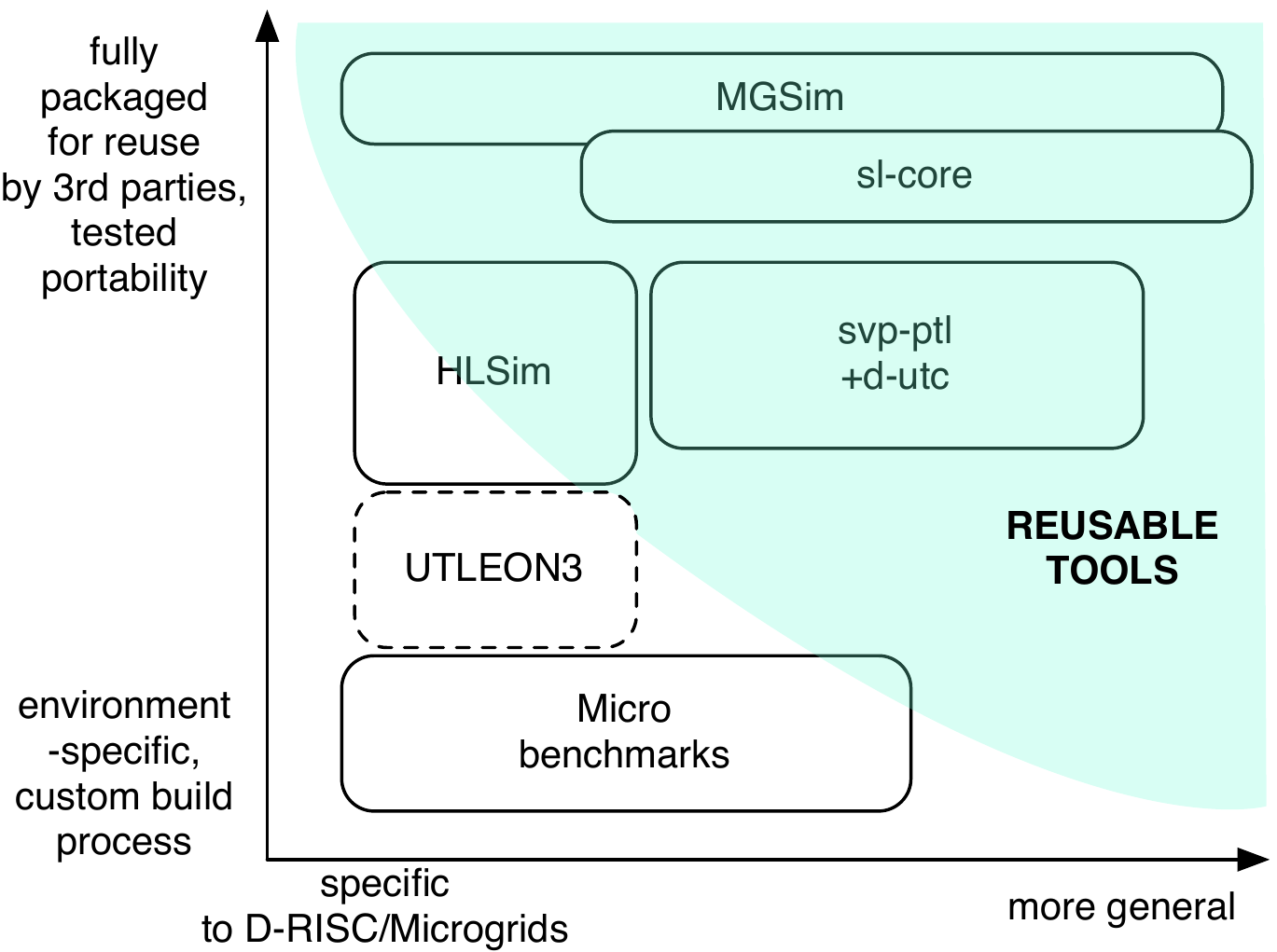}
\caption{Technology products as of 2012.}\label{fig:tech}
\end{figure}

Some of these tools are specific to the D-RISC/Microgrid architecture and are not applicable outside of this project, whereas
others could be reused by researchers that have never been exposed to D-RISC/Microgrids. On a different scale,
some of these tools have been explicitly packaged for reuse and tested for portability by 3rd party users, whereas
others are not readily reusable due to dependencies on the local research environment. I summarize how the tools map to
these two scales in \cref{fig:tech}.

\section{Conceptual separation of concerns}

The research around D-RISC/Microgrids has strongly promoted two
intellectual exercises and shaped a generation of researchers able to
converse fluently about the following two issue separations.

The first is the separation between concurrency and parallelism, \ie
the distinction between \emph{opportunity for parallelism} that can be
encoded in software by relaxing synchronization constraints, and the
\emph{actual simultaneity of execution at run-time} which depends on
resource duplication over space. 

This first separation is promoted/enabled by D-RISC/Microgrids by
promoting a TMU protocol which does not guarantee the availability of
parallel resources at run-time. Software that wishes to use the TMU
can only relax synchronization, \ie introduce concurrency, whereas
actual parallelism is introduced later, by the TMU at run-time,
depending on resource availability. Although this separation of
concerns can be promoted by other means, any researcher working on
D-RISC/Microgrids \emph{cannot avoid} acquiring a sharp consciousness
of these issues.

The second separation is between ``memory as storage'' and ``memory as
a synchronization mechanism.''  In commodity architectures, the only
interface between the individual core and its environment in a
multi-core chip is its memory interface. This implies that the same
memory interface is used for both loading and storing values to main
memory within individual threads, and for coordination of work between
cores. The latter, in particular, has historically mandated the
extension of memory systems with transactional mechanisms (bus
locking, compare-and-swap, test-and-set) which would otherwise be
unneeded.

As explained in~\cite[Chap.~7]{poss.12}, the proposed
D-RISC/Microgrids architecture separates\footnotemark{} the memory
network for data storage from a ``control network-on-chip'' in charge
of synchronizing and coordinating work between TMUs. This forces the
researchers writing software for the platform to realize that the data
structures for synchronization traditionally implemented in memory,
such as producer-consumer FIFOs, mutexes or semaphores, are really specific instances of
\emph{abstract synchronization services} whose behavior can be
obtained in other ways, possibly more cheaply and efficiently.

\footnotetext{The aim of this separation was to test whether a custom
  NoC can achieve cheaper and more efficient synchronization than a
  memory system, and to avoid the research overhead of developing a
  complex cache coherency protocol that also supports transactions.
  This aim was reached, insofar that a custom NoC is indeed cheaper
  than a comprehensive memory protocol, but it requires special
  support in software, \cf~\cite[Chap.~7]{poss.12}.}

These separation of concerns are not yet widely understood and commonly accepted in
the research community. The increased intellectual acuity
of the researchers ``educated'' by working on D-RISC/Microgrids
forms an advantage that can thus be considered a contribution of the enterprise.

\begin{summary}
\begin{itemize}
\item The research has produced interesting discussions
  that challenge some tacit assumptions of the research community, 
  experimental results that can be reused by future work, improvements
  to partner technologies and new simulation techniques.
\item Most of the software designed and implemented during the research
  can be reused by third parties, and not only for research
  directly related to D-RISC/Microgrids.
\item The intellectual framework educates practitioners to think about
  two general separations of concerns, namely concurrency \vs parallelism and
  using memory for storage \vs synchronization.
\end{itemize}

\end{summary}

\chapter{Individual architectural features}

Publications, posters and talks usually present the D-RISC core and
Microgrid clusters thereof as a single coherent technology made from
inter-dependent features. In reality, a gradual composition is
possible, as well as adding TMU-like features to other processor cores
than D-RISC.

I shortly present here my understanding of this
composition, which I have started to recognize while
writing~\cite[Chap.~3]{poss.12}. 

\section{Overview}

\begin{figure}[b!]
\centering
\includegraphics[width=.6\textwidth]{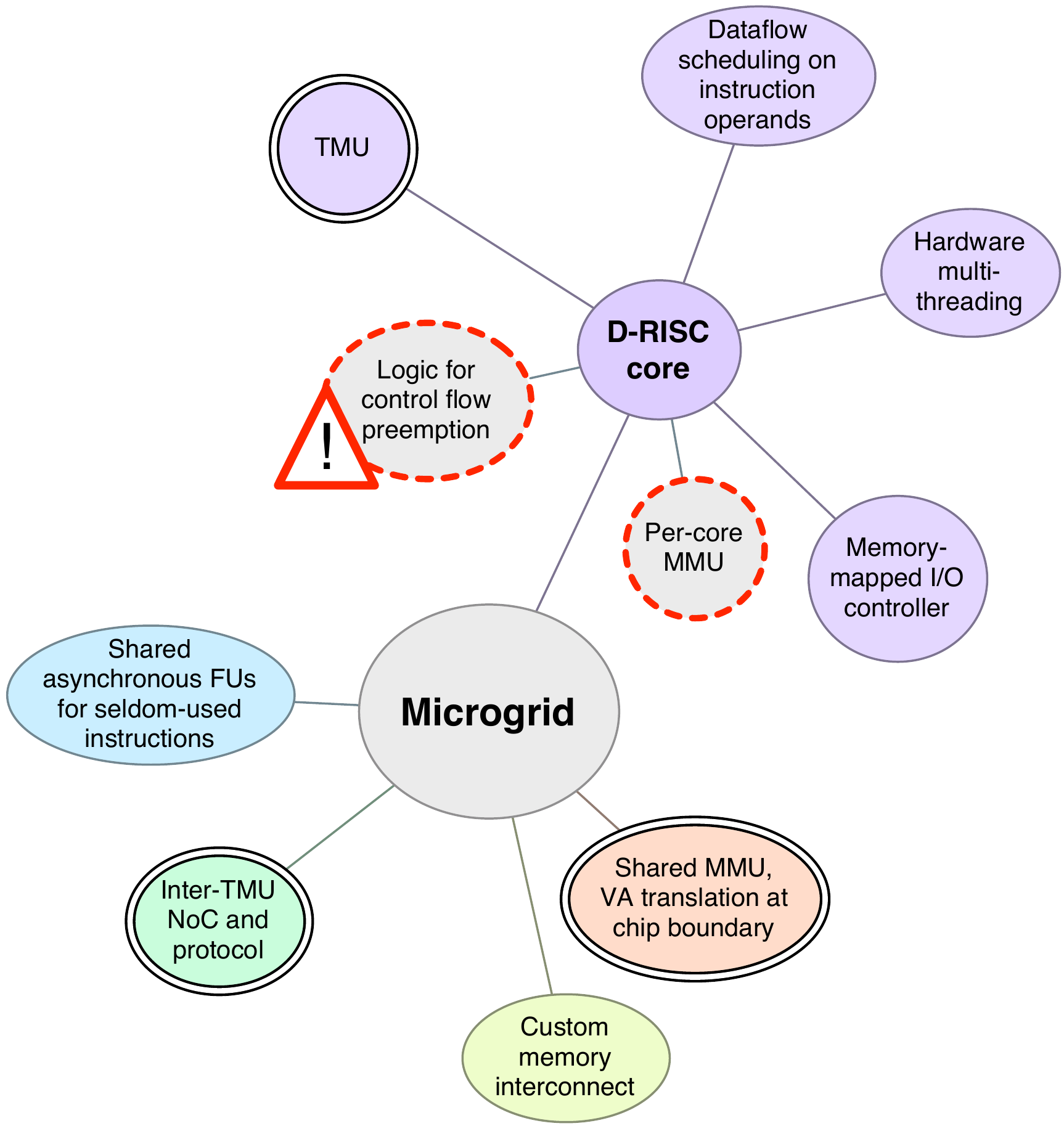}
\caption{Overview of the characteristic features of D-RISC/Microgrids.}\label{fig:features2}
\end{figure}

To start with, I summarize
the characteristic features in \cref{fig:features2}; in this
diagram I denote with a double edge the features not found
in other processors, and with a striped red edge those
features found in other processors but not on D-RISC/Microgrids. This diagram
exposes the composition of features in the design, as follows:

\begin{itemize}
\item the D-RISC core itself is a composition of the following features:
\begin{itemize}
\item a pretty conventional in-order, single-issue RISC pipeline,
\item a dataflow instruction scheduler that can execution instructions out-of-order while respecting data dependencies,
\item multiple hardware threads (separate program counters and registers),
\item a memory-mapped interface to an I/O subsystem,
\item a custom hardware Thread Management Unit (TMU) that manages logical tasks and maps and schedules them over hardware threads;
\end{itemize}
\item a Microgrid is a cluster of D-RISC cores together with:
\begin{itemize}
\item asynchronous functional units (FUs) between cores for seldom-used instructions;
\item a shared MMU that provides a single virtual address space to all cores;
\item a custom control network-on-chip to coordinate concurrency management between TMUs;
\item optionally, a custom memory interconnect.
\end{itemize}
\end{itemize}

Of these features, we can distinguish those that provide ``added
value'' compared to other processor architectures, namely the TMU,
dataflow scheduler and control NoC, from those that are ``unique'' and
make the design fundamentally incompatible with conventional wisdom, namely
the lack of support for preemption and the lack of per-core (and
per-hardware thread) MMU, which were discussed in \cref{sec:incompat}.

Remarkably, this overview alone reveals that the combination of
features found in D-RISC could be obtained by starting from an existing RISC
core design. The Tera MTA, for example, already features hardware
multithreading, a dataflow scheduler and memory-mapped I/O, and could
be \emph{extended} with a TMU. Similarly, Sun/Oracle's Niagara T4
cores already feature hardware multithreading, memory-mapped I/O and
out-of-order instruction execution (via reservation stations, which is
really a form of dataflow scheduling), and could thus be extended with
a TMU as well.

\section{Details}

\begin{figure}
\centering
\includegraphics[width=\textwidth]{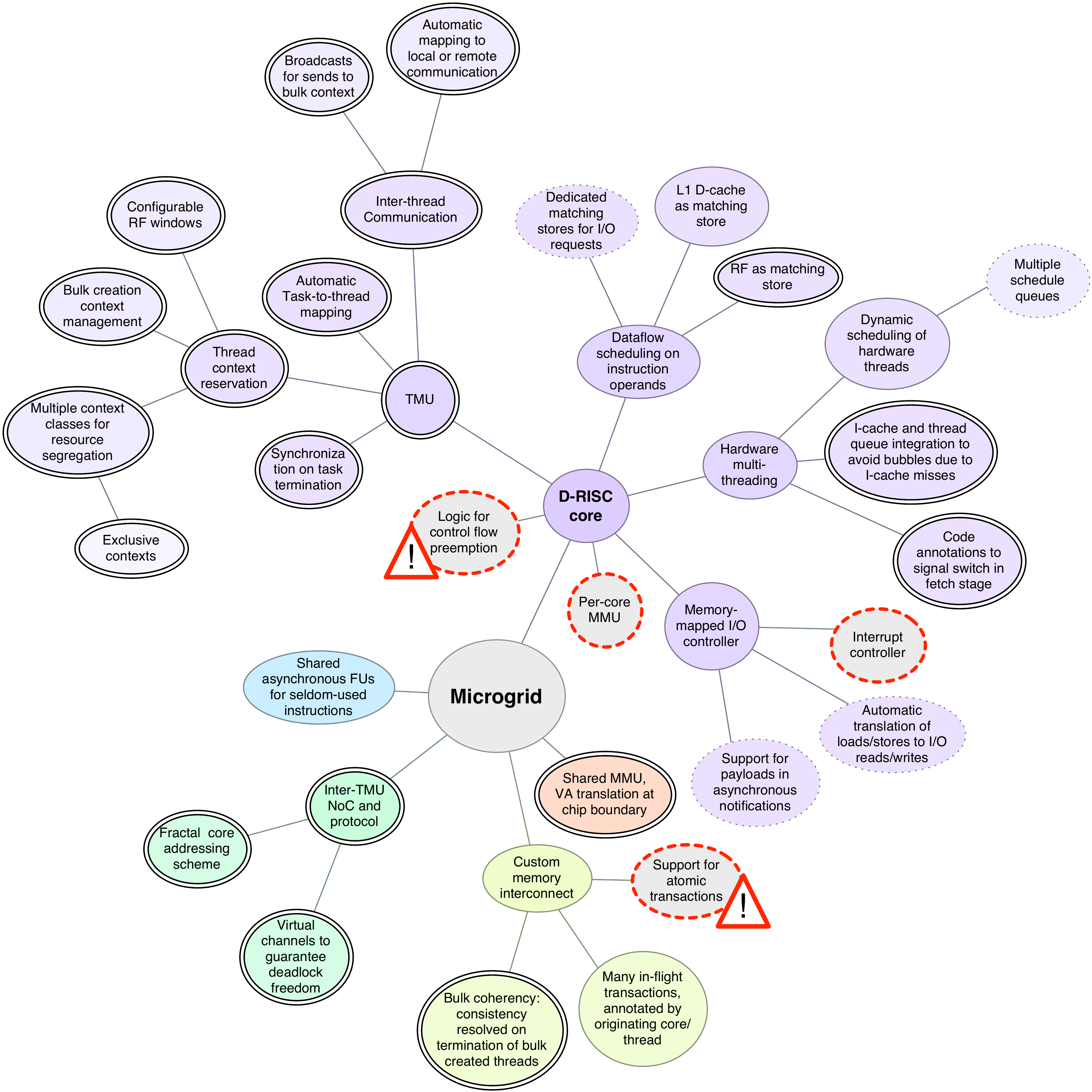}
\caption{Overview of the characteristic features of D-RISC/Microgrids (expanded).}\label{fig:features}
\end{figure}

To understand how D-RISC and Microgrids additionally benefit from re-implementing
its own ``version'' of features already found in existing processors,
it is useful to dig one level deeper, as illustrated in
\cref{fig:features}. 

(Again, I denote with a double edge the features
not found in other processors, and with a striped red edge those
features found in other processors but not on D-RISC/Microgrids. Features
with dotted borders are experimental, not yet ready for production. The
red triangles highlight the missing features that fundamentally hurt
software reuse.)

Besides highlighting the wealth of features of D-RISC's TMU and the inter-TMU NoC, this diagram draws
attention to the following points.

\subsection{Dataflow scheduling from the register file}

All dataflow schedulers rely on a \emph{matching store}, which retains
information about ``what to do'' when an operation has completed,
while the operation is ongoing. In conventional architectures,
matching stores are implemented:
\begin{itemize}
\item as part of the memory
sub-system, starting at the data cache (e.g. Tera MTA, Niagara T4+), to
  determine ``what to do when a memory operation completes,''
\item  via reservation stations next to functional units in out-of-order
execution (e.g. PowerPC), to determine ``what to do when a local operation completes.''
\end{itemize}

Like MTA and T4, D-RISC uses the L1 D-cache as matching store; but
contrary to conventional OoOE techniques it uses the main register
file instead of reservation stations for out-of-order execution of
instructions. 
This choice \emph{simplifies the circuits} that connects the pipeline,
functional units and the register file together. The trade-off is that
the number of in-flight instructions \emph{per thread} is limited to
the ISA register space, typically 31, whereas it can grow arbitrarily
with reservation stations.

This way, we can recognize that \emph{in the large design domain where
  out-of-order instruction execution is desirable, D-RISC is exploring
  the sub-domain where per-thread ILP can be traded off with simpler
  circuits.}

\subsection{Optimizations to hardware multithreading}

\Cref{fig:features} also highlight two features that optimize
hardware multithreading, regardless of the presence of the TMU
and of a dataflow scheduler.

The first is a tight integration of the schedule queue and the
I-cache.  As detailed in~\cite{lankamp.08} and
\cite[Sect.~3.2]{poss.12}, a a thread is not considered by the fetch
stage unless its code is already in the I-cache, so as to prevent
pipeline bubbles due to I-cache misses. This technique decouples code
fetching and thread scheduling and \emph{yields higher utilization of
  the pipeline overall}. I do not know whether this feature
is used by other hardware multithreaded core designs, but
I would be surprised if it were not.

The other feature is the ability to \emph{hint} the fetch stage of the
pipeline to switch preemptively to another thread after fetching an
instruction, when that instruction is \emph{likely to cause a pipeline
  bubble at a later stage}. This can be used for any instruction that
reads the result of a previously issued long-latency instruction. For
example, in ``ld r1 $\leftarrow$ [x]; add r2 $\leftarrow$ r1+r1'' the ``add''
instruction would be annotated, so that the fetch stage preemptively
switches to another thread: if the ``add'' suspends due to a missing
input (\eg due to a cache miss on the previous ``ld'' instruction), no
other instruction from the same thread needs to be flushed from the
pipeline. This feature also \emph{yields higher utilization of the
  pipeline overall}, to my knowledge is not present in other
processors, but could be possibly be added to them.

To this date, switch hints are implemented in D-RISC by
\emph{interleaving} hint bits with the instruction stream, which in
turns requires a custom assembler to generate the binary program
code. However, other implementations are possible, \cf
\cite[Sect.~4.4]{poss.12}.

\subsection{Bulk coherency in the shared memory network}

As soon as the designer of a chip architecture combines multiple cores
with memory-level parallelism and still wishes
to expose a ``shared memory'' interface to programmers,
a \emph{coherency protocol} must be designed to ensure
that the memory updates performed by one core on one part
of the memory become eventually visible to other cores
connected to other parts of the memory.

(Note that this problem disappears if all cores are physically
connected to the same memory or shared cache with a bus.)
 
\emph{Necessarily}, a coherency protocol implies communication across
the memory network to propagate the data ``placed into the memory'' by
store requests, to the point where it may be needed by subsequent load
requests.  Also, since the memory network cannot ``predict the
future,'' it must make a decision to effect this propagation
preemptively at some granularity: either after each individual store,
or while evicting cache lines, or by allowing the program code to
``grab exclusivity'' for a range of addresses over the entire system (write-invalidate).

In the Microgrid design, knowledge from the TMU about the program's
structure is used to implement an optimization to coherency: when a
bulk of tasks are created together (a feature of the TMU), the TMU
informs the cache network that it can wait to propagate the stores performed by
that bulk until all tasks in it have terminated, or until when a task creates
a sub-task. This ``makes sense,'' \ie it is valid, because D-RISC's programming
model specifies that stores are only visible to other tasks after a
task terminates or to the sub-tasks it subsequently creates. The
benefit of this optimization is a \emph{reduced number of
  coherency-related communication events in the memory network} for some workloads.

To my knowledge, this optimization opportunity is also exploited in
SIMD/SPMD accelerators, and in general it could be readily
considered in any design where high-level information about the
clustering of software operations is visible to the hardware.

\section{TMU reusability}

The purpose and consequence of decomposing the D-RISC design is to
isolate its TMU and recognize that the TMU is really a hardware
accelerator for system management functions that would be otherwise
realized in software. It is actually possible to describe the TMU as
an extension to any generic RISC core, even a core that does not
offer the other features of D-RISC:

\begin{itemize}
\item without hardware multithreading, the TMU would be constrained to
  schedule logical tasks over a single hardware thread. This would
  restrict the amount of instruction-level parallelism (because the
  maximum number of in-flight instructions is restricted by ISA
  register window), but would still save up the cost of branches and
  increments to implement repetition, and thus accelerate loops;
\item without out-of-order execution (either dataflow scheduling or
  via other means), instructions that control the TMU would cause the
  processor to wait until the TMU operation has completed. This may
  imply large waiting times for ``complex'' operations, for example
  allocating a group of cores on another part of the chip. It would
  also mandate the use of interrupts to signal asynchronous
  completion, for example the termination of a task, but would still
  save up the overhead of doing the thread management entirely in a
  software operating system.
\end{itemize}

\begin{summary}
\begin{itemize}
\item The D-RISC core combines features found in other processors, such as a RISC pipeline
  and hardware multithreading, with custom features (\eg its TMU) and optimizations
  to the conventional features (\eg switch annotations for the HMT scheduler).
\item Some architectural optimizations found in D-RISC/Microgrids could be reused 
  with other processors, for example switch annotations and bulk coherency in the memory network.
\item The key feature of D-RISC/Microgrids, namely its TMU and inter-TMU control NoC, 
  does not depend on the other features specific to D-RISC and could be potentially reused with other processors.
\end{itemize}
\end{summary}

\chapter{Follow-up strategies}

\Cref{part:decons} has highlighted shortcomings in the methodology and
obstacles to further progress. This analysis raised the question of
how to move forward from there \emph{differently}, so as to
avoid these shortcomings and obstacles. This chapter presents my view on this question, articulated in two directions: first,
what would constitute ``sane approaches'' for new projects (\cref{sec:new}); then
how to ``fix'' or ``improve'' ongoing projects/research (\cref{sec:ongoing}).

\section{Possible strategies for new investments}\label{sec:new}

\paragraph{Exploit.}

Apply the technology produced
so far to other uses than research.

A successful application so far has been education: with only
a minor but regular maintenance effort, the simulation tools
can provide support in architecture and compiler courses
for the coming 5-10 years.
However, given the processor is unable to support any C code that requires a ``hosted''
environment (or other
languages whose RTS is written using hosted C), applications in industry will be limited to small
embedded systems.

Possibly, with only a minor effort investment, an ad-hoc form of preemption
and per-core MMU can be added to a simulation model and obtain a
limited compatibility with software frameworks. Without
significant research, this would yield
sub-efficient (non-competitive) performance, but the gained
compatibility might be sufficient to activate further
external interest in the work.

\paragraph{Salvage and open.}

Extract individual features from the D-RISC/Microgrids
design and evaluate them as extensions
of existing processors.

Small ``first steps'' in this direction can be made
by starting with the switch annotations and the coupling
of the fetch stage with the I-cache. These features
seem readily applicable to the Niagara architecture
and the latest ARM multithreaded cores.
A more significant project would be to extract
the TMU and offer it as a reusable accelerator
component where the processor designer
can choose which TMU feature are activated. For example,
the features related to bulk creation/synchronization
or multi-core resource management may not be always relevant,
and a designer should not need to pay the price of
their integration if they end up not being used.

Conversely, the D-RISC core stripped of hardware multithreading and
its TMU could be offered as a SoC building block, marketing its dataflow
scheduler as a lightweight implementation of out-of-order execution. For
this block to be moderately competitive, a branch predictor
may be proposed as an option.

\paragraph{Distill and reincarnate.}

From the perspective of theoretical computer science, the
 D-RISC/Microgrids enterprise has raised two questions that may warrant
a wealth of further fundamental research.

The first was opened on purpose: \emph{how does the cost intuition of
  programmers evolve when complex operating system services are
  available at nearly the cost of basic arithmetic?} This is one of the key
questions that the TMU was designed to answer. The \emph{desired} answer was
originally: ``once programmers are comfortable about the costs of
concurrency, they would use concurrency everywhere and obtain parallel
speedups at every level.'' This particular answer was not obtained by
the research so far, but it may well be that other interesting answers can be
obtained instead.

Further effort in this direction could be bootstrapped as follows.
First, get acquainted with a software community already comfortable
using concurrency without too much assumptions about hardware.
Haskell and Erlang programmers are interesting candidates. Then,
observe and inventory which specific patterns of concurrency they
already use, and those they are striving to implement. Then re-design
a custom TMU that accelerates their favorite language run-time
system. Then, demonstrate the net effect on existing programs, and
document how the programmers modify their software over time to take
advantage of this accelerator. 

I discovered the other fundamental question while dismantling
the ``SVP model'' proposed in~\cite{jesshope.08.apc} and used
subsequently in the period 2008-2011. SVP has captured, using its
``places,'' the notion that a group of processors should be considered
as a \emph{single}, \emph{fungible resource} that can be allocated
dynamically from the computing environment and sub-partitioned
dynamically using abstract operators such as those implemented by
D-RISC's TMU. The designers of SVP then claimed that ``places are the
fundamental currency of computing'' and that their abstract operators
were ``general-purpose,'' \ie sufficiently general to carry out any computation. 

As I discuss in~\cite[Chap.~9\&12]{poss.12}, I believe this particular claim is invalid,
because a ``general'' model of computing resource should offer and
define memory and means for I/O as well, which SVP
places do not. However, studying SVP raises the complement question:
is it possible to \emph{extend a general model of computation with a
  cost model that uses entire virtual parallel computers with multiple
  cores and multiple memories as the basic resource unit}? A strategy
for exploring this question would 
probably benefit from starting with an inherently concurrent model, which
Turing and queue machines are not. The Actor model~\cite{agha.85} and
Milner's $\pi$-calculus~\cite{milner.92,milner.92.2} may be more suitable candidates, as their intuitive
implementations have well-understood operational semantics already. 

Further effort in this direction could be bootstrapped as follows. First,
select a technology which already uses virtualizations of entire
parallel resources as a basic building block. Modern Unix systems
and VM hypervisors are candidates. Then formalize its
basic concurrency operations (\eg fork, wait in Unix) in the conceptual
terminology of a general model. Then, based on this formalization and
expert knowledge of the actual behavior of the technology on parallel hardware,
design a cost algebra that is reasonably predictive. Then implement
a framework that visualizes and predicts cost for existing applications
using that technology. Use the interest gained in this way to attract
funding on the fundamental question.

\section{Possible strategies for ongoing projects}\label{sec:ongoing}

\subsection{Partnership with industry: 150k€ at stake}

An industry partner has recently funded some initial research effort to add
priority scheduling to D-RISC's thread scheduler and to explore fault
detection and recovery. Initial results suggest an opportunity to fund
further development effort in that direction, with the understanding
that the partner can use the benefits of the technology in their embedded
aeronautics controllers, which already use space-hardened custom SPARC
cores, in a 1-core or 2-core configuration.

Here the two strategies ``exploit'' and ``salvage and open'' described
above are applicable.

For the ``exploit'' strategy, the partner would need to fund
simultaneously a rewrite of the D-RISC specification in a language
suitable for both simulation and synthesis, so as to avoid maintaining
two source bases over time, and an extension of the current D-RISC
design to support preemption and resource reclamation, as much as
required by the partner's software.

For the ``salvage and open'' strategy, the partner could simply fund 
a rewrite of D-RISC's HMT scheduler and the subset
of D-RISC's TMU that is sufficient for the partner's software as an extension 
of the partner's favorite/desired existing processor core.

\subsection{Ongoing PhD theses: 400k€ at stake}

Both my peers who already defended a doctoral thesis founded on D-RISC and
Microgrids~\cite{bernard.10,vantol.13}, and myself, have been assailed during
our defenses with variation of the following:
\begin{itemize}
\item ``why did you choose this platform?'' 
\item ``what makes this platform especially attractive?''
\item ``why is your evaluation by software applications so poor?''
\end{itemize}

As answers, all three of us formulated variations of ``I was told this
platform was general enough and/or had great potential when I started,
and only later I recognized some of the obstacles, but I did my
part nonetheless. And look, by the way, I found some nice answers to
side research questions of my own, not initially phrased in the
D-RISC/Microgrids enterprise!''

Meanwhile, our unspoken thought was: ``I trusted my supervisor this
was the right place to start my PhD study and obtain the scientific
merit needed to graduate successfully, and as a beginner scientist I
did not have yet the critical acuity to recognize our shared
methodological shortcomings. But everyone can make mistakes, and
should be forgiven for them. After all, my PhD defense committee finds
me worthy of a doctorate, so it couldn't be as bad as it looks.''

In principle, the currently ongoing PhD research projects could
be concluded on the same note, and numerous new projects started
with the understanding they will conclude similarly. 

In practice however, as I am sitting next to them and entertain close
social contact, I feel dishonest letting my peers employ this strategy: given
I now understand the shortcomings, is it fair to let my peers struggle
with the large friction to academic publication and peer acceptance
caused by our communal continued use of a flawed approach? The risk
is great also that they recognize this friction but feel the obstacle
is insurmountable, or worse, that this realization engenders distrust
against the potentials of further research in the area.

Here, unfortunately, I do not have the experience sufficient to
guarantee better outcomes with an alternate strategy with any
confidence. The essence of any sane approach, to me, would be to
retrospectively \emph{reverse-engineer properly formulated research
  questions} that happen to be suitably answered by the work effectively performed,
independently from the initial initiative. This question should then
be phrased as generally as possible so that it does not hinge on the
specifics of D-RISC/Microgrids. Only in a second phase, subsequently
propose the current implementation of D-RISC/Microgrids as a case
study.  To illustrate, I list some possible rephrasings in
\cref{tab:foo}. In nearly all cases, I think it would be useful to
acknowledge early on that the restrictions described in
\cref{sec:incompat} are arbitrary, and seek actively means to overcome
them to gain access to more software benchmarks. This may even imply partial uses of the ``salvage and open''
strategy described earlier.

\begin{table}
\begin{tabular}{p{.35\textwidth}cp{.55\textwidth}}
Initial impulse & & Reverse-engineered research questions \\
\hline
how to build a D-RISC TMU? & $\rightarrow$ & what are the costs/benefits of
accelerating OS functions for thread management with a hardware unit? \\
 & $\rightarrow$ & what insights about how the hw/sw interface
 influences programming language semantics, are gained while building a TMU? \\
how to build a D-RISC/Microgrids simulator? & $\rightarrow$  & what simulation
 framework would be suitable for research in micro-architecture design while
 keeping simulation performance high enough to run significant multi-core workloads? \\
& $\rightarrow$ & to which level of accuracy can a model in this framework simulate
  the behavior of a hardware implementation? \\
how to improve D-RISC's memory performance? & $\rightarrow$  & what are the costs/benefits
  of modifying memory interfaces and protocols to increase the latency tolerance abilities of cores that use HMT and/or dataflow schedulers? \\
& $\rightarrow$  & what are the quantitative benefits of exploiting the concurrency awareness available
  in hardware in memory protocols? \\
how to implement priority scheduling in D-RISC? & $\rightarrow$  & what are the cost/benefits
 of extending a HMT scheduler with priorities? \\
how to implement fault tolerance in D-RISC/Microgrids? & $\rightarrow$  & what are the costs/benefits
  of exploiting the concurrency awareness available in hardware in fault tolerance protocols? \\
& $\rightarrow$ & is it possible to abstract fault tolerance to a general computing model equipped
with a resource/cost model?
\end{tabular}
\caption{Example reverse-engineering of research questions.}\label{tab:foo}
\end{table}

\begin{summary}
\begin{itemize}
\item I can see three follow-up strategies for new investments around
  D-RISC/Microgrids: exploitation, \ie apply the technology produced
  so far to other uses than research; salvaging and opening the
  technology, \ie extracting individual features from the
  D-RISC/Microgrids design and evaluating them as extensions of
  existing processors; and distillation of the main ideas in the realm
  of fundamental computer science.
\item Ongoing research towards doctoral theses should be careful
  to rephrase research questions in the light of our recent shared
 understanding of the project's issues.
\end{itemize}
\end{summary}

\chapter{Conclusion}

In traditional academic research projects, the abstract and general
questions receive most attention, and technology and engineering
``happen'' as a by-product. In contrast, the D-RISC/Microgrids project
was primarily a technology and engineering enterprise, with some
occasional and incidental scientific output.

My opinion is that further work in this direction faces two fundamental problems.

Firstly, a continued focus on engineering makes the project increasingly
  difficult to host in an academic institution and impedes the growth
  of an academic network.

  The product of the current and past effort is made of chip
  blueprints, simulation software, ancillary programming tools,
  education materials and demonstration tools. Unfortunately, the
  metrics used in academia to reward scientific effort are
  peer-reviewed academic publications, conference attendance, invited
  talks and lectures, successfully defended doctoral theses, etc. This
  mismatch implies that the work has become extraordinarily difficult to
  defend in academic communities. Moreover, any team member expecting
  to receive an academic training from this project risks facing a
  strong sense of disconnect between expectations and reality that may drive
  them away. This is detrimental to the growth of a network of
  supporting researchers around the project.

Secondly, the lack of connections with related work, especially
  a continued disregard for software compatibility, constitutes a serious
  management issue that threatens the project.

  This disconnect has not always been an issue. In general, at
  the start of a new line of research in computer architecture, compatibility
  can be readily sacrificed to simplify the research environment 
  and quickly obtain preliminary evaluation results using simple, ad-hoc
  experiments.  Moreover, ten years ago when the research strategy
  was being shaped, there did not yet exist any pervasive software culture for
  multi-core programming and software interfaces to concurrency management.
  In this context, a new, immature approach was simply competing with a host of other equally
  new, immature approaches. But this context has thus evolved, and
  the research risks facing irrelevance if the circumstancial changes in context and expectations are not addressed soon.

The question then remains: what to do now? For this, I have detailed
in \cref{sec:new} three possible strategies for new investments which
I know are viable from the current status of the research and would
address the two problems identified above.

One is to \emph{exploit}: take the shortest practical route to
maximize visibility of the current results and apply the technology. I
am currently driving exploitation towards academia, using the produced
tools for education in chip architecture and code generation. I am
seeking support from undergraduate students to design a minimal but
working form of exception handling and system-level compatibility with
existing software. I am also keeping ready to partner with industry to
work on exploitation projects that do not require further design.
Another is to \emph{salvage and open}: bring the technology apart and
offer its most salient bits and pieces as reusable components, able
to ground partnerships for follow-up joint research projects using
existing platforms and processor core designs. I may be interested
to support work in this direction, but not to drive the work myself.
The third is \emph{distill and reincarnate}: extract the underlying
fundamental research questions that are still relevant in this day and
age, and create a new research direction to explore. I have started
some preliminary work in this direction myself already.

An incidental, more personal but more fundamental question in the
bigger picture is whether any of these strategies is favorable to the
development of a researcher's career in the current academic
institution where the project is currently hosted.  According to my
hierarchy, the answer is currently: ``not likely.'' I may try to
convince them otherwise by sublimating the work somehow, but personal
circumstances may prevent my long-term dedication to D-RISC/Microgrids
in favor of more aligned research topics instead.






\newcommand{\etalchar}[1]{#1} 
\addcontentsline{toc}{chapter}{References}
\bibliographystyle{is-alphaurl} \bibliography{doc}

\end{document}